\documentclass[useAMS,twocolumn,usenatbib]{mn2e}

\usepackage{graphicx}
\usepackage{amssymb}
\usepackage[fleqn]{amsmath}
\usepackage{color}

\long\def\symbolfootnote[#1]#2{\begingroup%
\def\thefootnote{\fnsymbol{footnote}}\footnote[#1]{#2}\endgroup} 

\def\kpc{{\,\mathrm{kpc}}}
\def\kms{{\,\mathrm{km\,s^{-1}}}}
\def\Msun{{\,\mathrm{M}_\odot}}
\def\vlos{{v_{||}}}
\def\vesc{{v_\mathrm{esc}}}
\def\vmin{{v_\mathrm{min}}}
\def\pos{{\boldsymbol{x}}}
\def\intd{{\mathrm{d}}}
\def\rp{{r_\mathrm{peri}}}
\def\ra{{r_\mathrm{apo}}}
\def\ecc{{\epsilon}}

\title[Fast moving stars in SDSS] {On the Run: mapping the escape speed across the Galaxy with SDSS}
\author[Williams, Belokurov, Casey \& Evans]{Angus A. Williams$^1$,
  Vasily Belokurov$^1$, Andrew R. Casey$^1$ \& N. Wyn Evans$^1$
  \medskip
  \\$^1$Institute of Astronomy, University of Cambridge, Madingley Road, Cambridge, CB3 0HA, UK}
\begin{document}

\date{Accepted  Received ; in original form }

\pagerange{\pageref{firstpage}--\pageref{lastpage}} \pubyear{2016}

\maketitle

\label{firstpage}

\begin{abstract}
We measure the variation of the escape speed of the Galaxy across a range of $\sim 40\kpc$ in Galactocentric radius. 
The local escape speed is found to be $521^{+46}_{-30}\kms$, in good agreement with other studies. 
We find that this has already fallen to $379^{+34}_{-28}\kms$ at a radius of $50\kpc$. 
Through measuring the escape speed and its variation, we obtain constraints on the Galactic potential as a whole. 
In particular, the gradient in the escape speed with radius suggests that the total mass contained within $50\kpc$ is $29^{+7}_{-5}\times10^{10}\Msun$, implying a relatively light dark halo for the Milky Way. 
Our method represents a novel way of estimating the mass of the Galaxy, and has very different systematics to more commonly used models of tracers, which are more sensitive to the central parts of the halo velocity distributions. 
Using our inference on the escape speed, we then investigate the orbits of high--speed Milky Way dwarf galaxies. 
For each dwarf we consider, we predict small pericenter radii and large orbital eccentricities. 
This naturally explains the large observed ellipticities of two of the dwarfs, which are likely to have been heavily disrupted as they passed through pericenter.
\end{abstract}

\begin{keywords}
Galaxy: halo -- galaxies: kinematics and dynamics -- dark matter
\end{keywords}

\section{Introduction}

The fastest moving stars have long been a subject of fascination and speculation. 
\citet{Vi56} recounts a discussion between Kopff and Eddington on horse-racing, an enthusiasm of the latter. 
Kopff stated he was not interested, because of course one horse will always run faster than another.  
``But why'', retorted Eddington. ``When one star moves faster, you are very interested!'' 
The fastest moving stars are intriguing both because of the processes that accelerated them and because they are probes of their environment.

The local escape speed $\vesc$ is a measure of the depth of the potential well at the solar position $\Phi(R_\odot)$. 
Therefore, the velocities of the speediest stars passing through the solar neighbourhood can in principle provide information on the enclosed mass. 
There were early investigations on the local escape speed by \citet{Ca81} and \citet{Al82}, who suggested values $\sim 450\kms$.
The analysis methods were extended and systematized by \citet{Le90}, who emphasised the importance of modelling the shape of the tail of the velocity distribution. 
They provided a number of physically-inspired models and concluded that $\vesc$ lay between $450$ and $650 \kms$ with 90\% confidence. 
The recent spectroscopic surveys of the Galaxy have stimulated a surge of activity, notably by \cite{Sm07} and \cite{Pi14} (hereafter S07 and P14).
These authors used data from the RAVE spectroscopic survey, which provided an abundance of information on the kinematics of local stars. 
P14, who defined the escape speed as the minimum speed needed to reach three virial radii, concluded that it was $533^{+54}_{-41} \kms$ (90\% confidence), albeit on the basis of a small sample.

There are some obvious drawbacks. First, there is no guarantee that the high velocity tail of the distribution function is actually occupied all the way up to the escape speed. 
This may mean that the velocity of the fastest moving star is an under-estimate of the true escape speed. 
Second, the method is sensitive to interlopers or contaminants, which may be unrepresentative of a smooth, relaxed stellar population. 
This includes stars in the process of leaving the Galaxy, such as hypervelocity stars ejected by interaction of binaries with black holes~\citep[e.g.,][]{Br15,Bou16}. 
Stars may also be unbound from the Milky Way but nonetheless bound to the Local Group. 
Although no such stars are known, the phenomenon is familiar to us through the intergalactic stars identified in nearby clusters like Fornax~\citep{Th97}. 
Third, and perhaps most awkwardly, the spatial distribution of the highest energy stars is set by the stochastic patterns of cosmic accretion. 
Therefore, the concept of a smooth velocity distribution may be a fiction at the highest energies, even at locations in the inner galaxy.

These disadvantages are offset by a number of assets. First, only the high velocity tail of the distribution function (DF) need be modelled, as opposed to its entirety. 
This is clearly a massive simplification that sidesteps much of the complexity in building DFs. 
Second, the method provides a nice counterpoint to other ways of measuring the potential (or equivalently the mass) of the Galaxy. 
For example, methods based on the Jeans equations use the first and second moments of the distribution, and so are controlled by the main bulk rather than the tail. 
And, third, future prospects are bright, with huge new data sets of radial velocities from spectroscopic surveys and of proper motions from the Gaia astrometric satellite becoming available.

All previous work has focussed on measuring the escape speed locally. 
This is because samples of high velocity stars have been small (sometimes minute, for example S07 used just 16 stars, whilst P14 relied on 86 stars) and concentrated in the solar neighbourhood.
In this paper, we present the first measurements of the escape speed throughout the Galaxy using a variety of tracers -- main-sequence turn-off stars (MSTOs), blue horizontal branch stars (BHBs) and K-giants -- extracted from the Sloan Digital Sky Survey.
Although the MSTOs are located at heliocentric distances within $\sim 3$ kpc, the BHBs and K-giants in our sample extend out to Galactoctocentric radii of $\sim 50$ kpc, providing much greater reach.

Section~\ref{sec:method} describes the likelihood function and models for the tail of the velocity distribution, primarily following the formalism established by \citet{Le90}. 
We discuss our tracers and distance estimators in Section~\ref{sec:samples}. 
This includes the cuts used to build the samples of MSTOs, BHBs and K giants, as well as to extract the high velocity stars. 
The choice of potential is given in Section \ref{sec:pot} and enables us to broaden the discussion of escape speed into enclosed mass and circular speed. 
Priors and numerical implementation are described in Section~\ref{sec:pannm}, whilst Section~\ref{sec:res} presents our results, and discusses their implications for three fast moving Galactic satellite galaxies (Bootes III, Triangulum II and Hercules).

\section{Method}

\label{sec:method}

\cite{Le90} proposed that the velocity distribution of high-speed stars is a power law of the form
\begin{equation}
  p(v)\propto\begin{cases}
    (\vesc - v)^k & \text{if $\vmin \leq v<\vesc$},\\
    0 & \text{otherwise}.
  \end{cases}
  \label{eq:lt90}
\end{equation}
$\vesc$ is the escape speed and $\vmin$ is a cut-off, such that $p(v < \vmin)$ begins to deviate from a power law. 
Since the model depends only on the speed $v$, the full velocity distribution function is implicitly isotropic. 
As radial velocities are measured far more precisely than transverse velocities, it is useful to marginalise the above model over proper motion. 
This gives
\begin{equation}
  p(v)\propto\begin{cases}
    (\vesc - |\vlos|)^{k+1} & \text{if $\vmin \leq \vlos <\vesc$},\\
    0 & \text{otherwise},
  \end{cases}
\end{equation}
where $\vlos$ is Galactocentric radial velocity. 
Since we are only interested in the absolute value of $\vlos$ in this work, we shall now refer to $|\vlos|$ as $\vlos$ for brevity.

This model has been applied to data several times, most notably by S07 and P14 using data from the {\sc RAVE} survey \citep{Ko13}. 
Both studies used small samples of stars ($<100$) close to the sun and found $\vesc \sim 530\kms$. 
We seek to extend their work by constraining the escape speed of the Galaxy at a variety of locations. 
We do this by parameterising the escape speed as a function of position $\pos$, so that
\begin{equation}
  p(\vlos \,|\, \pos) = \begin{cases}
    C\cdot(\vesc(\pos) - |\vlos|)^{k+1} & \text{if $\vmin \leq \vlos <\vesc(\pos)$},\\
    0 & \text{otherwise},
  \end{cases}
  \label{eq:model}
\end{equation}
where $C$ is a location-dependent normalisation factor, given by
\begin{equation}
C = \dfrac{k+2}{(\vesc(\pos) - \vmin)^{k+2}}.
\label{eq:norm}
\end{equation}
By far the largest source of uncertainty in our analysis is in the distance to each star. 
Consequently, we consider the uncertainty in the radial velocity, longitude and latitude to be negligible. 
Our likelihood function should therefore be the probability of a radial velocity, given Galactic coordinates and an imperfect inference on the distance to the star. 
Writing $\pos = (\ell, b, s)$, where $(\ell,b)$ are Galactic longitude and latitude, respectively, and $s$ is the measured line of sight distance to the star, we then have
\begin{equation}
p(\vlos \,|\, \ell, b, s) = \int p(\vlos \,| \,\ell, b, s' )p(s'\,|\,s)\, \intd s'.
\label{eq:crux}
\end{equation}
$p(s'\,|\,s)$ is the probability that $s'$ is the true distance to the star given our imperfect inference $s$. 
Finally, we also include a Gaussian outlier model for possibly unbound stars
\begin{equation}
p_\mathrm{out}(\vlos) = \dfrac{A}{\sqrt{2\,\mathrm{\pi}\,\sigma^2}}\exp \dfrac{-\vlos^2}{2\sigma^2},
\end{equation}
where we fix $\sigma = 1000\kms$, and $A$ is the normalisation of the Gaussian over the interval $\left[v_\mathrm{min},\infty\right]$. 
We then introduce a free parameter $f$ for the fraction of outliers, so that the overall likelihood function is
\begin{equation}
p_\mathrm{tot}(\vlos \,|\, \ell, b, s) = (1-f)\,p(\vlos \,|\, \ell, b, s) + f\,p_\mathrm{out}(\vlos).
\end{equation}
This Equation represents the likelihood that we will use for the remainder of the paper, while making specific choices for $\vesc(\pos)$. 
In practice, we compute the RHS of Equation (\ref{eq:crux}) using Monte-Carlo integration \citep[e.g.][]{Ev16,Bo16}, so that
\begin{equation}
p(\vlos \,|\, \ell, b, s) \simeq \dfrac{1}{N}\sum\limits_{n=1}^{N} p(\vlos \,| \,\ell, b, s_n ),
\end{equation}
where each of the $s_n$ is drawn from $p(s'\,|\,s)$.

\section{Sample selection}

\label{sec:samples}

\begin{figure*}
\includegraphics[width=2\columnwidth]{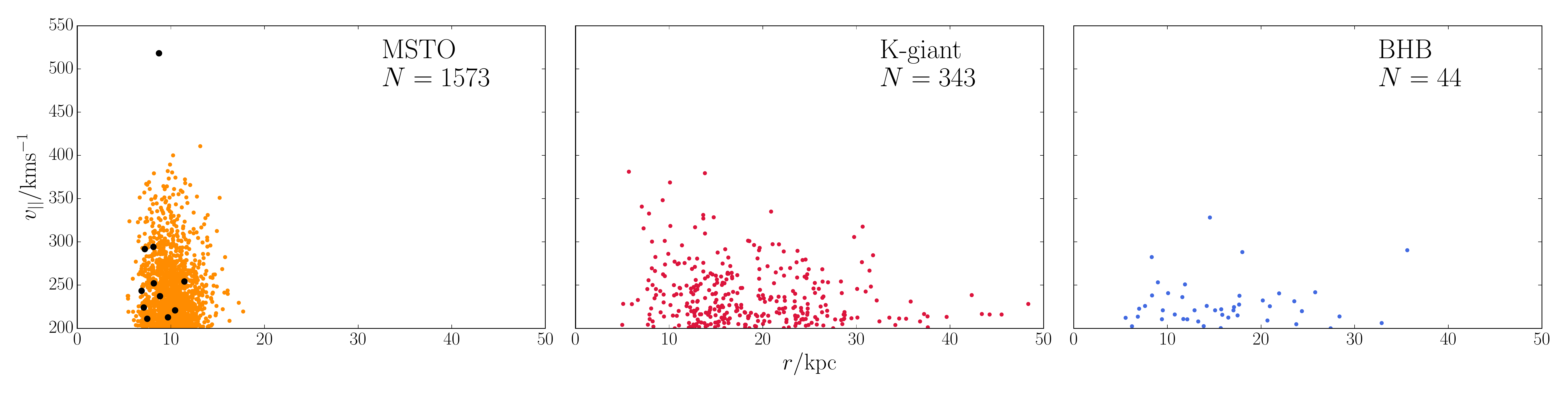}\\
\caption{The distribution in Galactocentric radius and $\vlos$ of each of our tracer samples. 
The MSTO stars probe a smaller range in radius, but are numerous, whereas the K-giants probe a much larger distance range, but are fewer in number.
The BHBs probe a similar range to the K-giants, but form a far smaller sample.
The black points in the MSTO panel highlight the 11 objects that are discussed in Section \ref{sec:res}.}
\label{fig:data}
\end{figure*}

In order to measure the variation in the escape speed with position in the Galaxy, we require a set of halo tracers that spans a sufficiently large volume, and has measured radial velocities and distances. 
To obtain such a sample, we use the SDSS 9th data release \citep{Ah12}. 
We choose to utilise three distinct sets of tracers: main sequence turnoff (MSTO), K-giant and blue horizontal branch (BHB) stars. 
MSTO stars are numerous, and are mostly observed at distances $\sim 3\kpc$. 
These stars allow us to constrain the local escape speed, and its variation relatively close to the sun. 
K-giants and BHBs are fewer in number, but are bright and have been observed at distances $\sim 50\kpc$ from the sun, pushing the spatial extent of our sample to a range of $\sim 40\kpc$ in Galactocentric radius. 
Before selecting our high-speed sample of stars, we first constructed a `mother sample' for each class of tracer using a series of cuts. 
In addition to the cuts described below, we also removed stars with latitudes $|b| < 20^\circ$ in order to remove possible disc contaminants, and stars with radii $r>50\kpc$. 
We compute Galactocentric radii by assuming a solar radius $R_\odot = 8.5\kpc$. 
The full {\sc SQL} queries used are given in Appendix \ref{sec:sql}.

\subsection{MSTO sample}

To extract the mother sample of MSTO stars, we start by selecting in the de-reddened colour-magnitude box
\begin{gather}
0.2 < (g-r)_0 < 0.6, \nonumber \\
14.5 < r/\mathrm{mag} < 20,
\end{gather}
where the $r$-band extinction $A_r<0.3$. We then make cuts on spectroscopic parameters, so that
\begin{gather}
4500 < T_{\mathrm{eff}}/\mathrm{K} < 8000, \nonumber \\
3.5 < \log g < 4, \nonumber \\
-4 < \mathrm{\lbrack Fe/H \rbrack} < -0.9.
\end{gather}
This gives us a sample of metal-poor MSTO stars. 
We also make cuts to ensure high-quality photometry and radial velocity measurements for the sample. 
To compute distances to these stars, we estimate the absolute magnitude in the $r$-band using the prescription derived by \cite{Iv08}. 
With $x=(g-i)_0$, we have
\begin{gather}
\delta M_r = 4.5 - 1.11\,\lbrack \mathrm{Fe/H} \rbrack -0.18\,\lbrack \mathrm{Fe/H} \rbrack^2, \nonumber \\
\begin{split}
M_{r0} = -5.06 + 14.32\,x - 12.{}&97\,x^2 + 6.127\,x^3 \nonumber \\
                    {}&- 1.267\,x^4 + 0.0967\,x^5,
\end{split}\\
M_r = M_{r0} + \delta M_r.
\end{gather}
Using these formulae, we compute a point estimate of the distance to each star and remove objects with implied heliocentric distances $>$ 15 kpc in order to remove objects with spurious absolute magnitude estimates.
This finally leaves us with a mother sample of 22071 MSTO tracers. 
In order to draw samples from $p(s' \,|\, s)$ (Equation \ref{eq:crux}), we take the quoted uncertainties on $g$, $r$, $i$ and $\lbrack \mathrm{Fe/H} \rbrack$ from the SDSS pipelines and assume that they are normally distributed and uncorrelated. 
We then draw Monte-Carlo samples from each of these distributions and compute the distance for each sample.

\subsection{BHB sample}

To obtain a clean sample of BHBs, we first select in the de-reddened colour-colour box
\begin{gather}
-0.25 < (g-r)_0 < 0, \nonumber \\
0.9 < (u-g)_0 < 1.4,
\end{gather}
which was used by \cite{De11}. 
There will still be significant contamination from high surface-gravity blue stragglers within this box, so we then make the spectroscopic cuts
\begin{gather}
-2 < \lbrack \mathrm{Fe/H} \rbrack < -1, \nonumber \\
3 < \log g < 3.5, \\
8300 < T_\mathrm{eff}/\mathrm{K} < 9300, \nonumber
\end{gather}
giving us 1039 BHBs in total. 
We estimate the absolute $g$-band magnitudes, and hence distances, of the BHBs using the relation derived by \cite{De11}
\begin{equation}
\begin{split}
M_g = 0.434 - 0.169\,(g{}&-r)_0+2.319\,(g-r)_0^2  \\
         {}&+20.449\,(g-r)_0^3+94.617\,(g-r)_0^4.
\end{split}
\end{equation}
We then use the same Monte-Carlo strategy as for the MSTO stars to sample from the distance uncertainties.

\subsection{K-giant sample}

Rather than gathering a sample of K-giants directly from SDSS, as we did for our other two tracer samples, we use the catalog constructed by \cite{Xu14}, taken from SEGUE. 
The SEGUE survey \citep{Ya09} explicitly targeted K-giants in three different categories: ``I-colour K-giants'', ``red K-giants'' and ``proper-motion K-giants''. 
All three categories satisfy the photometric constraints
\begin{gather}
0.5 < (g-r)_0 < 1.3, \nonumber \\
0.5 < (u-g)_0 < 2.5,
\end{gather}
and have measured proper motions $<11 \,\mathrm{mas\,yr^{-1}}$. 
Each target category then populates distinct regions of the $(u-g)--(g-r)$ plane. 
The precise details of the targetting criteria are given in \cite{Ya09}. On top of the SEGUE selection, \citeauthor{Xu14} then make further cuts such that
\begin{gather}
\log g < 3.5, \nonumber \\
E(B-V) < 0.25 \,\mathrm{mag}. \nonumber
\end{gather}
Besides our cuts on latitude and Galactocentric radius, we make one further cut on this sample to ensure a clean halo population, namely
\begin{equation}
\lbrack \mathrm{Fe/H} \rbrack < -0.9.
\end{equation}
After applying our latitude and metallicity cuts, there remain 5160 K-giants in our mother sample. 
\citeauthor{Xu14} compute posterior samples for the distance moduli, $\mu$, of the K-giants. 
Their table provides the maximum a posteriori value of $\mu$ for each star, percentiles of the posterior samples, and a summary of the uncertainty, $\delta \mu$. 
In order to compute distance samples for the K-giants, we simply assume that the uncertainty in the distance modulus is Gaussian, with a standard deviation of $\delta\mu$.

\subsection{High-speed sample}

Given our three mother samples, we now seek to select members of the high-speed tail of the velocity distribution. 
In order to do this, we must make a decision for the value of $v_\mathrm{min}$ in Equation (\ref{eq:model}). 
S07 did not have access to measurements of stellar parameters for their sample, and adopted a very high threshold of $v_\mathrm{min} = 300\kms$ in order to avoid contamination from the Galactic disc. 
P14 used proper motion measurements to remove stars with disc-like signatures of rotation (although in practice this cut operates very similarly to a metallicity threshold, see their Figure 9), and then introduced a cut of $v_\mathrm{min}=200\kms$. 
Our cuts in latitude and metallicity serve to remove disc contaminants, which is much less of a concern when using more distant SDSS tracers than it is for samples of RAVE stars. 
Since we expect very little disc contamination, our choice of $v_\mathrm{min}$ now depends only on the range of velocities for which we believe the power law model in Equation (\ref{eq:model}) is valid.

P14 performed a detailed analysis of the simulation suite of \citet{Sc09} in order to make an educated choice for $v_\mathrm{min}$, and found that the distribution function of $\vlos$ in the simulations does not significantly deviate from a power-law when $\vlos>150\kms$. 
In our work, we must consider the fact that our cut must be appropriate across a range of locations. 
Since the vast majority of our sample are further from the Galactic center than the sun, and physical reasoning suggests that the escape speed cannot increase as a function of radius, then a cut of $200\kms$ should guarantee that the power-law model of Equation (\ref{eq:model}) is appropriate at the locations of all the stars in our study. 
Consequently, we set $v_\mathrm{min}=200\kms$.  
Note that this cut is applied to Galactocentric radial velocities, and so the motion of the sun must be removed beforehand. 
For this, we assume a local standard of rest $v_\mathrm{LSR} = 240\kms$ and a solar peculiar motion $\left(U_\odot,V_\odot,W_\odot\right) = \left(11.1,12.24,7.25\right)\kms$ \citep{Sc10}.
Once this cut is applied, there remain 1573 MSTO stars, 343 K-giants and 44 BHBs. 
The distributions of each of our final tracer samples in the $r-\vlos$ plane is shown in Figure \ref{fig:data}. 
We note that our sample of 1960 stars represents a factor of $>100$ increase from the 16 stars used by S07 and a factor $>20$ compared to the 86 used by P14.

\section{Choice of $\vesc(\pos)$}

\label{sec:pot}

Our primary model for the escape speed is a spherically--symmetric power law model (SPL)
\begin{equation}
\vesc(r) = \vesc(R_\odot)\,\left(\dfrac{r}{R_\odot}\right)^{-\alpha/2},
\end{equation}
where $r$ is Galactocentric radius and $0 \leq \alpha \leq 1$, on physical grounds.  
We normalise the model to the escape speed at the solar radius. 
The gravitational potential $\Phi$ and escape speed are related by
\begin{equation}
\vesc = \sqrt{-2\Phi},
\end{equation}
meaning that the SPL model corresponds to a gravitational potential of the form
\begin{equation}
\Phi(r) = -\dfrac{v_\mathrm{circ}(R_\odot)^2}{\alpha}\,\left(\dfrac{r}{R_\odot}\right)^{-\alpha}.
\label{eq:splgrav}
\end{equation}
Thus, unlike other studies that measure only the local escape speed, we can translate our measurements of $\vesc(R_\odot)$ and $\alpha$ into inference of the circular speed of the potential, and the mass enclosed within spherical shells. 

We will also report on results when two other models are used. 
The first is a simple generalisation of the SPL, a power law in elliptical radius (EPL)
\begin{equation}
\vesc(r_q) = \vesc(R_\odot)\,\left(\dfrac{r_q}{R_\odot}\right)^{-\beta/2},
\end{equation}
with $r_q = \sqrt{R^2 + z^2/q^2}$, where $R$ and $z$ are the usual cylindrical coordinates, and $0 \leq q \leq \infty$. 
If the escape speed falls off more rapidly with height above the Galactic plane than it does in the radial direction, then $q<1$ (oblate), whereas larger escape speeds high above the plane suggest $q>1$ (prolate). 
This model corresponds to a gravitational potential of the same form as Equation (\ref{eq:splgrav}), but with elliptical radius $r_q$ replacing spherical radius $r$.

The final model we consider is the truncated flat rotation curve (TF) model \citep[see][]{Gi14,Wi99}. 
We primarily investigate this model in order to facilitate a direct comparison with the mass inference of \citet{Gi14}.
The model has a flat rotation curve of amplitude $v_0$ in the inner parts, which then starts to decline as a power-law $\gamma/2$ around a scale radius $r_s$
\begin{equation}
v_0^2(r) = \dfrac{v_0^2\,r_s^\gamma}{(r_s^2 + r^2)^{\gamma/2}}.
\end{equation}
The form of the escape speed for this model cannot be written in terms of elementary functions, so we omit it here.

\section{Priors and numerical implementation}

\label{sec:pannm}

Having constructed our likelihood function and gathered our data, we now need to choose explicit priors on each of our model parameters. 
We split these parameters into two groups: global parameters, which are the model parameters that are independent of our choice for $\vesc(\pos)$, and potential parameters, which concern our explicit model of the escape speed.

\subsection{Global priors}
We allow a different power law slope in the velocity distribution of each tracer $\boldsymbol{k} = (k_\mathrm{MSTO},k_\mathrm{K-giant},k_\mathrm{BHB})$, and we fit for the outlier fraction, $f$.
These four parameters constitute our global model parameters. 
We anticipate that the three values of $k$ should be similar, since all three sets of tracers belong to the halo.
On the other hand, the MSTO sample spans a different radial range to the BHB and K-giant samples, and so any differences in our inferred $k$ values could suggest a radial variation in $k$.

S07 and P14 both used cosmological simulations in order to place informative priors on the value of $k$. 
S07 used a flat prior over the range $2.7 < k < 4.7$ and P14 adjusted this range to $2.3 < k < 3.7$. 
These priors were necessary for both analyses because of the limited number of stars in both the S07 and P14 samples. 
Given our much larger sample of stars, we opt for a much less informative prior on each $k$, which is flat in the range $0 < k < 10$. 
This loose prior will allow us to critically assess the similarity between cosmological simulations and the Galaxy. 
Our prior on $f$ is flat in the permissible range $0 \leq f \leq 1$.

\subsection{Potential priors}

\begin{figure*}
\includegraphics[width=2\columnwidth]{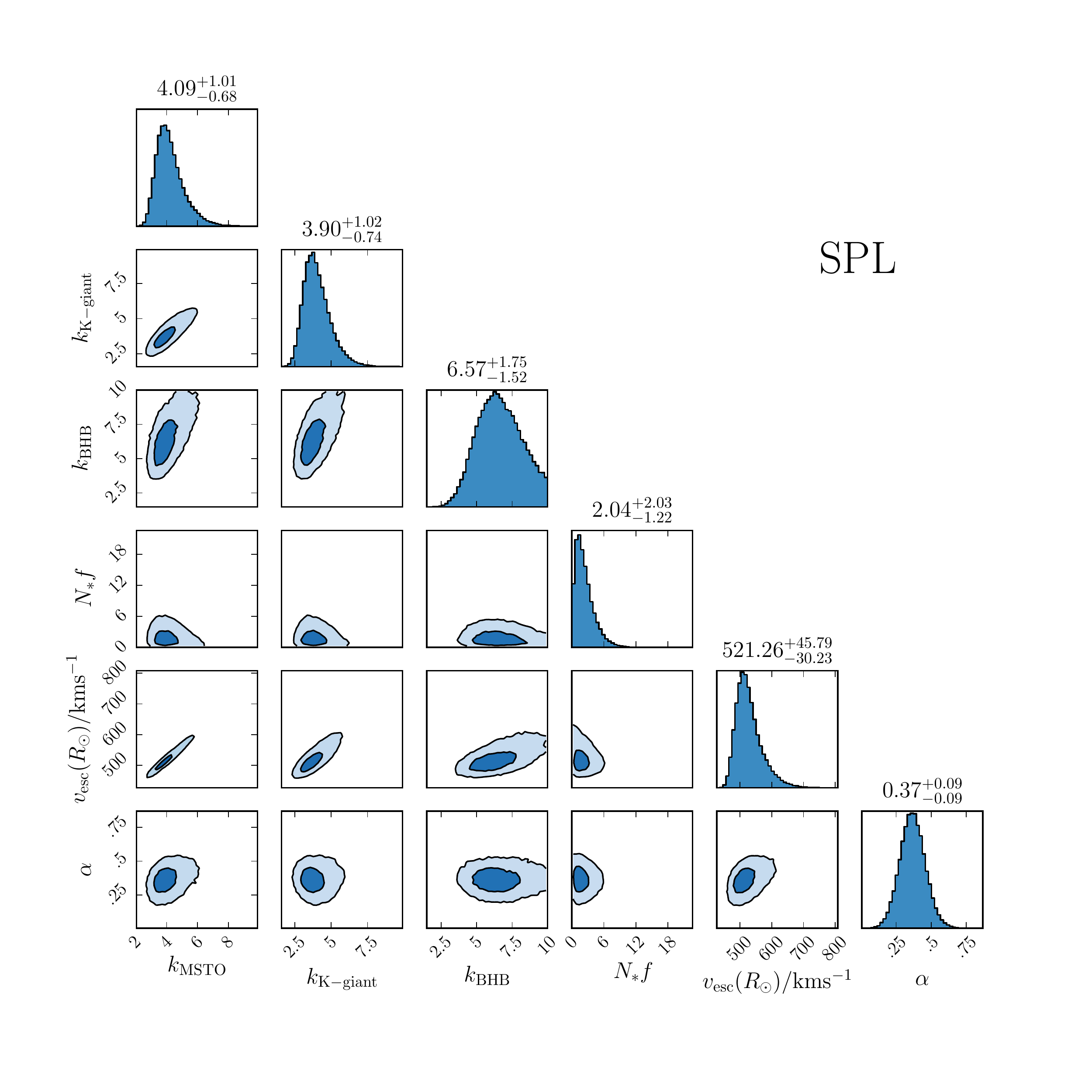}\\
\caption{One and two--dimensional projections of our MCMC samples for the SPL fit. 
The 68\% and 94\% credible intervals are shown in the 2D projections, and the median parameter values and uncertainties computed using the $\pm 34\%$ credible intervals of the 1D projections are shown above the 1D histograms. 
We multiply the outlier fraction $f$ by the number of stars in our sample, $N_*=1960$ in order to make the inferred value easier to interpret.
All of our parameters are well constrained, except $k_\mathrm{BHB}$, which is unsurprising given that this is the tracer sample with the smallest number of stars by an order of magnitude. 
Notable degeneracies are between $k_\mathrm{MSTO}$ and $k_\mathrm{K-giant}$, which are constrained to be close to equal, {} and between $\vesc(R_\odot)$ and $k_\mathrm{MSTO}$ and $k_\mathrm{K-giant}$. 
Larger values of $k$ suggest larger values of $\vesc(R_\odot)$, as expected. 
See text for discussion.}
\label{fig:SPL_corner}
\end{figure*}

We now detail the priors used on the parameters on each of the three forms for $\vesc(\pos)$.
The SPL model has two parameters: $\theta = (\vesc(R_\odot),\alpha)$. 
If the Galaxy is to possess a flat or declining rotation curve, as we would expect from physical reasoning, then
\begin{equation}
0 \leq \alpha \leq 1.
\label{eq:powerlawprior}
\end{equation}
Besides this basic physical constraint, we do not include any more information about the value of $\alpha$, so our prior is flat over the above range. 
Since $\vesc(R_\odot)$ is a positive definite scale parameter, we adopt a scale invariant Jeffreys prior
\begin{equation}
p\left(\vesc(R_\odot)\right) \propto 1/\vesc(R_\odot).
\label{eq:jeffreys}
\end{equation}
The EPL model is identical to the SPL model other than the inclusion of the axis ratio, $q$, so that $\theta = (\vesc(R_\odot),\,\beta,\,q)$. 
The priors we use for $\beta$ and $\vesc(R_\odot)$ are the same as in Equations (\ref{eq:powerlawprior}) and (\ref{eq:jeffreys}). For $q$, we follow \citet{Bo16} and use a prior
\begin{equation}
p(q) \propto \dfrac{1}{1+q^2},
\end{equation}
which places equal weight on oblate ($0\leq q <1$) and prolate ($q>1$) axis ratios.

Our final model, the TF model, has three parameters: $\theta = (v_0,\,r_s,\,\gamma)$. 
For the velocity normalisation $v_0$, we use a Jeffreys prior, limited so that $150 < v_0/\mathrm{kms^{-1}} < 400$. 
For the scale length, $r_s$, we use a prior that mirrors the result found by \citet{Gi14}, so that
\begin{equation}
\log p(r_s) \propto -\dfrac{1}{2}(r_s - \mu)^2 / \sigma^2,
\end{equation}
where $\mu=15\kpc$ and $\sigma=7\kpc$. 
We adopted this prior after experimentation with less informative priors, where we found that our model, in combination with these data, does not constrain $r_s$ (i.e., the posterior distribution on $r_s$ always resembles the prior).
The power law index $\gamma$ can take any value between 0 and 1, and hence we adopt the same prior as Equation (\ref{eq:powerlawprior}).

\subsection{Sampling method}
\label{sec:sampling_method}

Armed with our priors, we can now perform a Bayesian analysis. 
Our full parameter space is $\Theta =(\boldsymbol{k},f,\theta)$, which has 6 (SPL) or 7 (EPL and TF) dimensions. 
Bayes' theorem for our problem can be written in the following way
\begin{gather}
\begin{align}
p(\Theta \,| \,\mathrm{data}{}&) =  \nonumber \\
                    {}& \dfrac{p(\theta)\,p(\boldsymbol{k},f)\,\prod\limits_{j=1}^3
                    \prod\limits_{i=1}^{N_j}p_\mathrm{tot}(v_{||,i}^j \,|\, \ell_i^j,b_i^j,s_i^j,k_j,f,\theta)}{p(\mathrm{data})},
\end{align}
\label{eq:bayes}
\end{gather}
where the index $j$ refers to the tracer type (MSTO, K-giant or BHB). 
Since we do not seek to compare the evidence for our three models, we do not explicitly compute the denominator of Bayes' theorem. 
In order to constrain the model parameters, we use an MCMC approach.

In order to compute the likelihood, we evaluate the integral in Equation (\ref{eq:crux}) using 200 Monte-Carlo samples from the distance uncertainties of each star. 
We draw these samples once, and then use the same set for each posterior evaluation, thus avoiding random noise in the posterior \citep{Mc13}.
We then use the {\sc emcee} code \citep{emcee} to Monte-Carlo sample the posterior distribution. 
{\sc emcee} is a {\sc python} implementation of the affine--invariant ensemble sampling approach suggested by \citet{Go10}, where an ensemble of $N$ `walkers' is used, and the proposal distribution for a given walker is based on the current positions of $N/2$ of the remaining walkers. 
Here, we choose $N=80$.
We initialise the ensemble by randomly sampling from our prior distributions, and then evolve each walker for 5000 steps. 
We then inspect the trace plots in each dimension in order to prune our samples for burn-in, which typically requires $\sim 200$ steps. 
To assess the convergence of our chains, we first compute the acceptance fraction $a_f$.
For all the analyses presented in this paper, $0.3 < a_f < 0.5$ for the entire ensemble of walkers.
In addition, we compute the integrated autocorrelation time $\tau_f$ for each of our chains, which is the number of posterior evaluations required to produce independent samples.
We find $\tau_f \simeq 50$ in each dimension, so that our chains are run for $\sim 100$ autocorrelation times, which provides us with $\sim 8000$ independent samples from the posterior.

\section{Results}

\label{sec:res}

\begin{figure}
\includegraphics[width=\columnwidth]{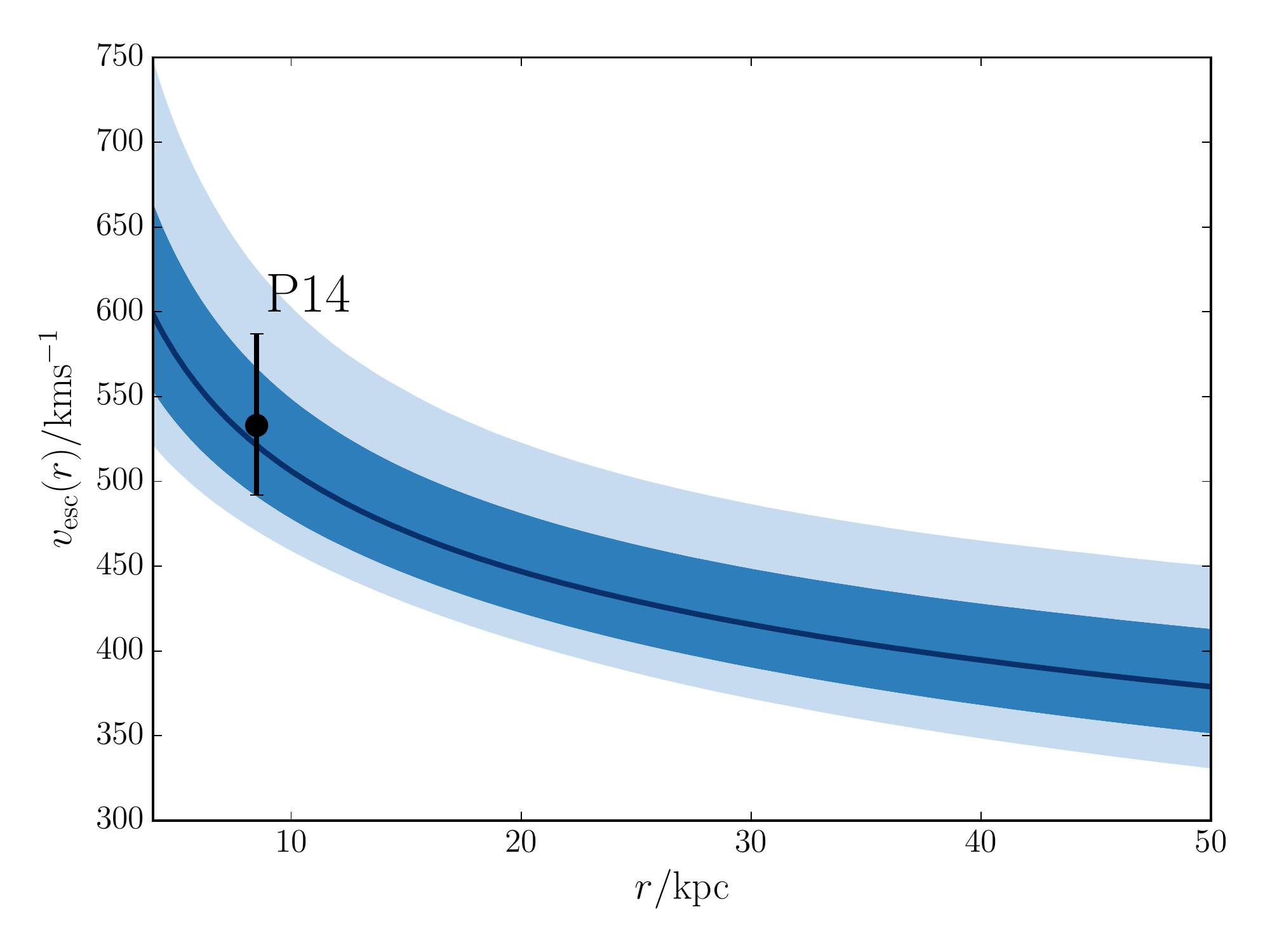}\\
\caption{Our inference on the escape speed as a function of Galactocentric radius. 
The median posterior result is shown as a {} dark blue line, and the 68\% (94\%) credible interval is a dark (light) blue band. 
The result using RAVE data from P14 and the associated 90\% credible interval is also shown, and is in good agreement with our inference. 
We measure a significant gradient in the escape speed, such that it has already fallen by $\sim 100\kms$ by  a radius of $30\kpc$.}
\label{fig:vesc_posterior}
\end{figure}

\begin{figure}
\includegraphics[width=\columnwidth]{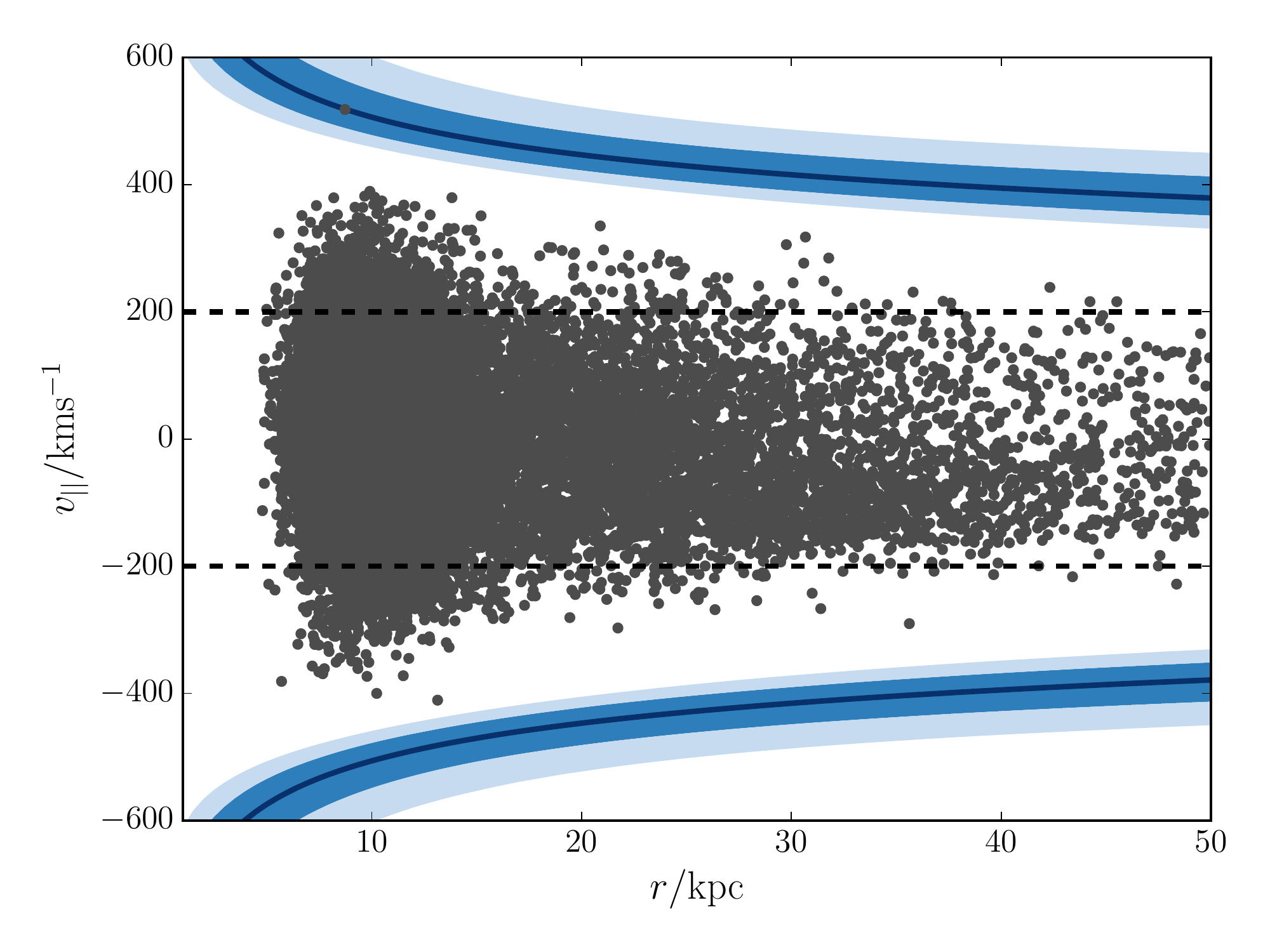}\\
\caption{The distribution of our mother samples of stars in the $r-\vlos$ plane, with horizontal dashed lines at $\vlos = \pm 200\kms$, our cut in radial velocity. 
The coloured bands are our inference on the escape speed as a function of radius.
The `spur' at negative radial velocities is from K-giants belonging to the Sagittarius stream. 
Note that the contamination in our high speed sample from these stars is negligible, since the maximum velocity that the stream centroid reaches is $\sim 150\kms$ \citep{Be14} with a dispersion of $\sim 20\kms$.}
\label{fig:stars_bound}
\end{figure}

Figure \ref{fig:SPL_corner} shows the one and two dimensional projections of the posterior samples of the six SPL model parameters, as well as the inferred median posterior values and uncertainties based on the $\pm 34\%$ credible intervals. 
All of the model parameters are well constrained, save for $k_\mathrm{BHB}$, which is unsurprising given that BHBs are by far the least numerous tracer in our sample.

Our results imply a local escape speed of $521^{+46}_{-30}\kms$, which is in good agreement with S07 and P14.  
We infer a power law index $\alpha = 0.37\pm0.09$, suggesting that the escape speed is falling rapidly as a function of radius.
The middle panel of Figure \ref{fig:data} is prophetic of this, because the edge of the K-giant distribution in the $r-\vlos$ plane is steep.
Figure \ref{fig:vesc_posterior} shows the run of $\vesc$ with radius implied by our inference, with associated 68\% and 94\% credible intervals, and the steep drop in the escape speed is clear.
For perspective, we also show the distribution of the mother samples of each tracer group in the $r-\vlos$ plane in Figure \ref{fig:stars_bound}.
The Milky Way loosens its grip on its inhabitants significantly: our model predicts that the local escape speed is $521^{+46}_{-30}\kms$, and by $50\kpc$ this has dropped to $379^{+34}_{-28}\kms$.

A priori, it is unclear what the value of $k$ should be. 
\citet{Le90} point out that violent relaxation would lead to $k=3/2$, whereas collisional relaxation gives $k=1$ \citep{Sp72}. 
S07 further showed that the Plummer and Hernquist spheres \citep{BT08} have $k=2.5$ and $k=3.5$, respectively. 
The simulations analysed by S07 and P14 both suggest $k\simeq3$. 
Clearly, there is a relatively large range of possible values.  
Due to small sample sizes in previous studies, $k$ has never been measured from data on the Milky Way. 
Given our significantly larger sample of stars, we are able to do this for the first time. 
The two tracer samples containing the most stars, MSTO and K-giants, both favour $k\simeq4\pm1$, which is in comfortable agreement with simulations.
These results also suggest that $k$ is not a strong function of position, given the rather different radial ranges probed by the MSTO and K-giant samples.
The inference on $k$ for the BHB sample is much weaker, and favours a slightly higher value. 
S07 points out that this is to be expected for small sample sizes. 
Nonetheless, the inference on $k_\mathrm{BHB}$ is not in significant tension with the hypothesis that $k$ is constant. 
Our results vindicate the choice of prior by S07, while the range used by P14 is a touch on the low side.

Figure \ref{fig:SPL_corner} shows a strong degeneracy between $k_\mathrm{MSTO}$ and $\vesc(R_\odot)$, which can be encoded by the empirical covariance matrix of the samples
\begin{equation}
\mathrm{Cov}\left(k_\mathrm{MSTO}\,,\,\vesc(R_\odot)\right) = \begin{bmatrix} 0.84 & 37\kms \\ 37\kms & 1713 \mathrm{km^2s^{-2}}\end{bmatrix}.
\end{equation}
This is to be expected, and is the reason why a narrow prior on $k$ was necessary in previous work. 
Figure 1 of P14 nicely demonstrates the appearance of this degeneracy for varying sample sizes. 
Fortunately, our sample is large enough to locate the maximum along the degeneracy. 
The same degeneracy is seen between $k_\mathrm{K-giant}$ and the local escape speed, though it is broader.
Note that this explains why our statistical uncertainty on the local escape speed is larger than that of of P14, who found $533^{+54}_{-41}\kms$ at 90\% confidence, compared to our 90\% credible interval of $521^{+88}_{-45}\kms$.
Our larger 95th percentile of $690\kms$ is a consequence of the degeneracy between $k$ and $\vesc(R_\odot)$: the 95th percentile of the posterior on $k_\mathrm{MSTO}$ is 6, which is considerably larger than the upper end of P14's prior.

The inferred outlier fraction is very small, $f \simeq 0.001$, but non-zero. 
This suggests that there are one or two outliers in our sample. 
Inspection of Figure \ref{fig:data} suggests one clear candidate: there is an MSTO star at $r\simeq10\kpc$, shown as a black point, with a measured line of sight velocity of $518.2\kms$, which is more than $100\kms$ larger than any other star at a comparable radius in our sample. 
Otherwise, there are no obvious outliers through visual inspection. 
As a check of this intuition, we calculated the outlier probability of each star in our sample as 
\begin{equation}
p(\mathrm{outlier}\,|\,\vlos,\ell,b,s) = \dfrac{\bar{f}\,p_\mathrm{out}(\vlos)}{\bar{f}\,p_\mathrm{out}(\vlos) + 
                                                      (1-\bar{f})\,p(\vlos | \ell,b,s,\bar{\theta},\bar{\boldsymbol{k}})},
\end{equation}
using the model parameters obtained by taking the median values of each of the one dimensional marginalised posterior distributions, $\bar{\Theta}$.  
The largest outlier probability is $>0.999$, and belongs to the object identified visually in Figure \ref{fig:data}. 
Otherwise, the largest outlier probability is $<0.01$, and so we conclude that this object is the only probable outlier in the sample. 
Having identified this outlier, we visually inspected its spectrum and image data from SDSS.
From the image data it is clear that this object is a galaxy, and has been misclassified by the spectroscopic pipeline of SDSS.
Having found a galaxy contaminant in our sample, we added a further constraint to our {\sc SQL} query that all of the MSTO targets should be morphologically classified as stars (as well as spectroscopically, as in our original query) in order to check for any other similar contaminants.
The high speed sample is reduced in size by 11 objects when this stricter constraint is applied: all 11 of these targets, including the galaxy outlier previously discussed, are shown as black points in Figure \ref{fig:data}.
We visually inspected the spectra and image data for all of these objects, and concluded that, other than the original outlier that we identified, they are likely partially blended stars.
Although this might mean that the photometry for some of these objects may be unrepresentative to some degree, visual inspection of their spectra suggests that they likely have reliable stellar parameters and radial velocities.
For this reason, we opted not to remove them from our sample, but we did re-run our analysis with these objects removed to check for inconsistencies.
We found that none of our conclusions change when these targets are removed from the sample: the comparison of the posterior distributions with and without these objects is shown in Appendix \ref{sec:morph}.
Later, when we compare our model to the data, we remove the bona-fide galaxy contaminant from the sample, but retain the other 10 objects.

\begin{figure}
\includegraphics[width=\columnwidth]{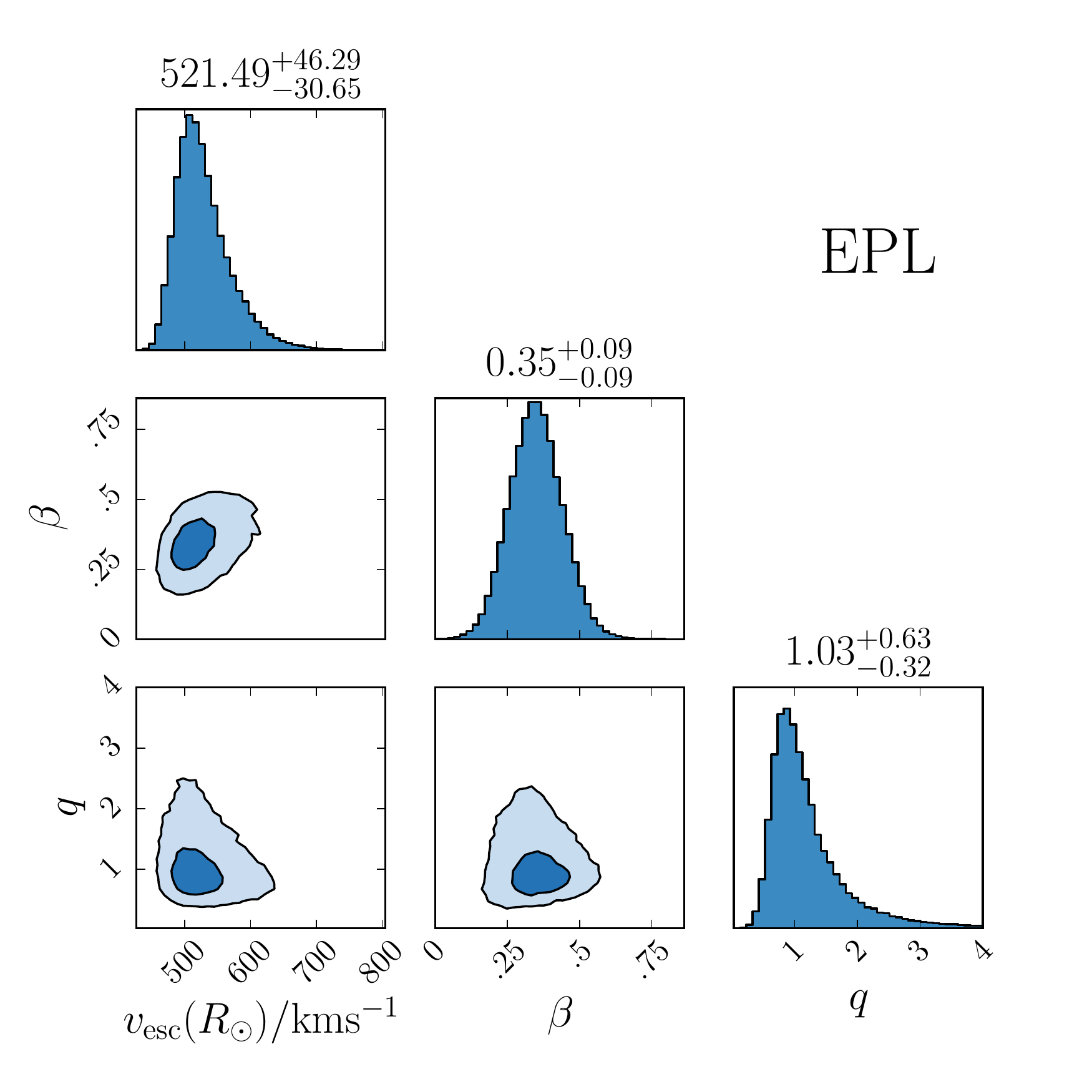}\\
\caption{Posterior distribution for the EPL model. 
The results are very similar to the SPL model, and the extra parameter $q$ is found to be $\sim 1$, suggesting that the Galactic potential is probably spherical, though our uncertainties are quite large.}
\label{fig:EPL_posterior}
\end{figure}

\begin{figure}
\includegraphics[width=\columnwidth]{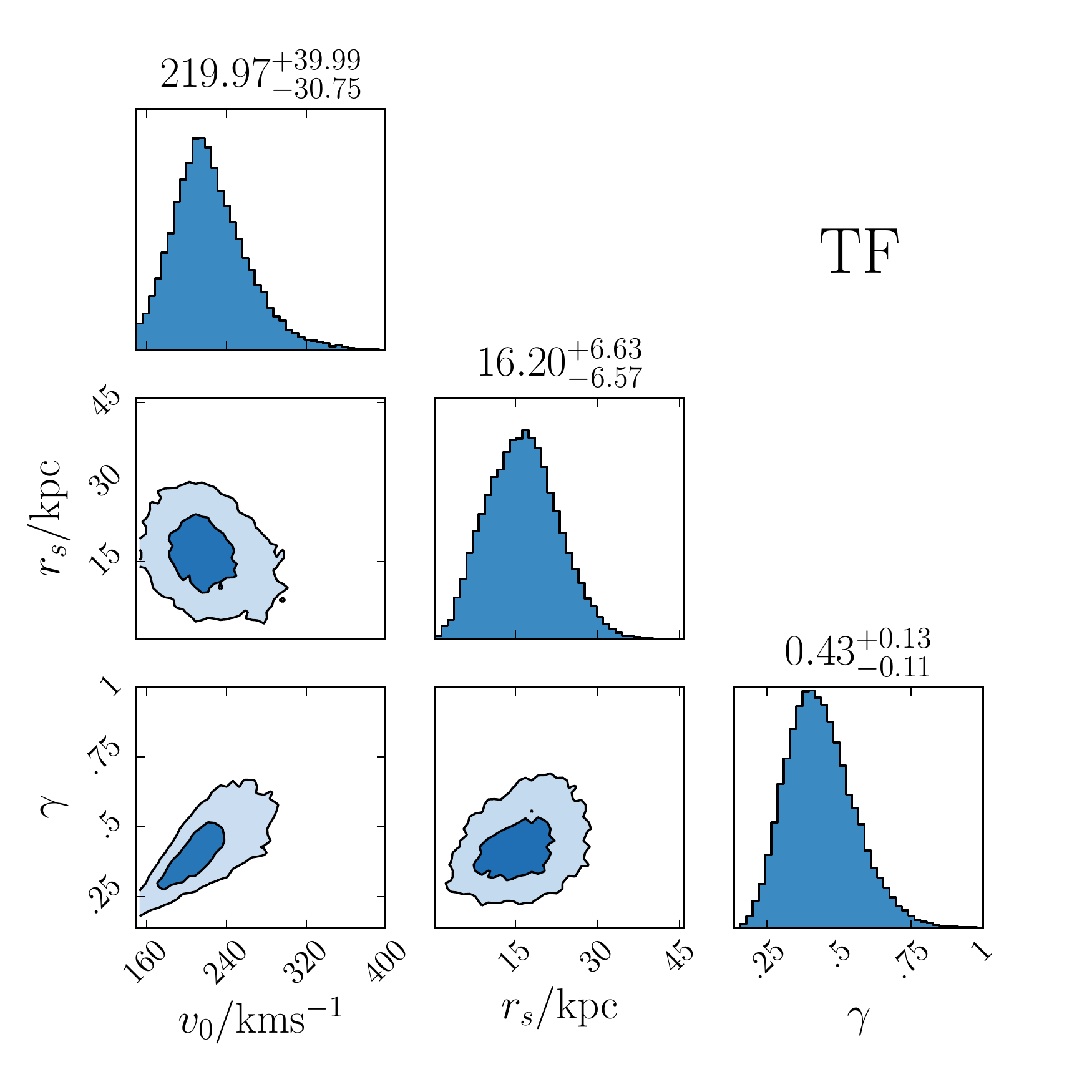}\\
\caption{Posterior for the TF model when the prior based on the inference of \citet{Gi14} is used for $r_s$. 
We infer a rotation curve amplitude of $220\kms$ and a power law decline of $\gamma \simeq 0.4$, both of which are entirely consistent with the results of \citet{Gi14}.}
\label{fig:TF_posterior}
\end{figure}

Figures \ref{fig:EPL_posterior} and \ref{fig:TF_posterior} show the results of the EPL and TF analyses. 
We removed the global parameters from the figures because the results were essentially identical to those found for the SPL model, without any interesting additional degeneracies. 
The addition of a halo axis ratio, $q$, leads to the conclusion that the escape speed falls at the same rate in all directions from the Galactic centre, with $q = 1.03^{+0.63}_{-0.32}$. 
This in turn implies that the Galactic potential is likely spherical, although our uncertainties are large: the data are compatible with $q=0.7 - 1.6$. 
Since $q$ corresponds to the flattening of the potential, this corresponds to a relatively wide range of flattenings in the dark matter halo. 
Hopefully, with more data, the method will provide a more useful constraint on the symmetries of the Galactic potential. 
The other two parameters of the EPL model are the same as those in the SPL model, and the inferred values and uncertainties are indistinguishable.

Our inference on the TF model suggests an inner rotation curve with amplitude $v_0 = 220^{+40}_{-31}\kms$, which starts to fall as $\gamma/2 = 0.22^{+0.07}_{-0.06}$ over a scale $r_s = 16^{+7}_{-7}\kms$. 
The posterior on the scale radius is no different from our prior, and so is uninteresting. 
However, having assumed similar inference on the scale radius as \citet{Gi14}, the amplitude and power law index are entirely consistent with their result, which we will discuss further in Section \ref{sec:mass}. 
When translated into an escape speed profile, the inference appears virtually identical to Figure \ref{fig:vesc_posterior}. 
This suggests that our inference is relatively robust against different parameterisations of the escape speed.

\subsection{The mass and rotation curve of the Milky Way}
\label{sec:mass}

Having mapped the escape speed across the Galaxy, we are now able to convert this measurement into a mass profile $M(r)$ and rotation curve $v_c(r)$ for the Galaxy. 
For a spherically symmetric escape speed profile, we have
\begin{eqnarray}
M(r) &=& \dfrac{-r^2\,\vesc}{G}\dfrac{\mathrm{d}\vesc}{\mathrm{d}r}, \label{eq:mass}\\
v_c(r) &=& \sqrt{-r\,\vesc\,\dfrac{\mathrm{d}\vesc}{\mathrm{d}r}}. \label{eq:circspeed}
\end{eqnarray}
Given our model of $\vesc(r)$, and our inference on its parameters, we can compute posterior distributions on $M(r)$ and $v_c(r)$ using these formulae.
For example, the local circular speed that we measure is
\begin{equation}
v_c(R_\odot) = 223^{+40}_{-34}\kms.
\end{equation}
Note that our inferences on the mass profile and rotation curve rest heavily on the assumption that the speed distribution of stars in the Galaxy truncates at $\sqrt{-2\Phi}$.
If the stars do not fill out to the this value, then our analysis will underestimate the depth of the potential well, which will lead to underestimates in the inferred mass profile and rotation curve of the Galaxy.
P14 showed that the inferred halo virial mass increased by 20\% if the escape speed was instead defined as $\sqrt{-2\left(\Phi(r) - \Phi(r_\mathrm{max})\right)}$, with $r_\mathrm{max} = 3R_\mathrm{vir} \sim 600\kpc$.
Since we do not attempt to track the Galaxy's mass out to such large radii, the possible bias incurred by effectively setting $r_\mathrm{max} = \infty$ will be significantly smaller than 20\%, and so we henceforth assume that the velocity distribution reaches $\sqrt{-2\Phi}$.
The value of $v_c(R_\odot)$ that we obtain while making this assumption is pleasingly aligned with a multitude of other methods, and provides us with confidence that systematic uncertainty caused by these considerations is unimportant relative to our statistical uncertainties.

It is worth noting that our method clearly possesses very different systematic uncertainties when compared to more common approaches in the literature.
Most dynamical models of halo tracers, like distribution function and Jeans analyses, are most sensitive to the central parts of the velocity distributions.
This is particularly true of Jeans analyses, which generally only model the first and second moments of the velocity distributions. 
Distribution functions satisfy the full collisionless Boltzmann equation, and therefore the entire infinite hierarchy of Jeans equations, but this generally comes at the cost of large systematic uncertainties that arise from the chosen form of the model \citep{Wa15}.  
Our approach moves the focus to the tail of the velocity distribution, and is therefore complimentary to other approaches.

Figure \ref{fig:mass} shows the mass and circular speed profiles implied by the SPL model, along with associated 68\% and 94\% credible regions. 
Our model predicts $M(50\kpc)=29.8^{+6.9}_{-5.2} \times 10^{10}\mathrm{M}_\odot$. 
For reference, we have also plotted the results from a selection of other studies. 
\citet[][X08]{Xu08}, \citet[][D12]{De12} and \citet[][WE15]{Wi15} all used samples of halo BHBs taken from SDSS. 
D12 and WE15 applied distribution function models to the data, and infer systematically higher masses than we do here, with $M(50\kpc) \simeq 45\times10^{10}\mathrm{M}_\odot$.  
Both are consistent with the 94\% credible interval of our inference, but there is a hint that there is a discrepancy between distribution function methods and the present approach.
X08, on the other hand, compared SDSS BHBs to cosmological simulations, and their result is comfortably in agreement with ours.

W99, like D12 and WE15, used a distribution function approach, but applied their method to globular clusters and dwarf galaxies. 
Their sample was small, with only 27 objects, and so their uncertainty is large. 
Their preferred mass of $M(50\kpc)=54^{+2}_{-36}\times 10^{10}\mathrm{M}_\odot$, like the other two distribution function approaches, is significantly larger than our result, although the large asymmetric uncertainty removes any possible tension.

The final study to which we compare is that of \citet[][G14]{Gi14}, who modelled the disruption of the Sagittarius stream. 
They exploited the fact that the apocentric precession of the stream should be sensitive to the details of the gravitational potential. 
Their inference produces very similar results to our work, with $M(50\kpc)=29\pm5\times 10^{10}\mathrm{M}_\odot$. 
We can make this comparison even more explicit because we have also estimated the parameters of the model they used in their analysis (TF). 
When we use the TF model, the mass enclosed is $M(50\kpc)=33^{+8}_{-6}\times 10^{10}\mathrm{M}_\odot$. 
The two analyses, though very different in detail, produce near identical results.

\subsection{The orbits of Milky Way dwarf galaxies}

\begin{figure}
\includegraphics[width=\columnwidth]{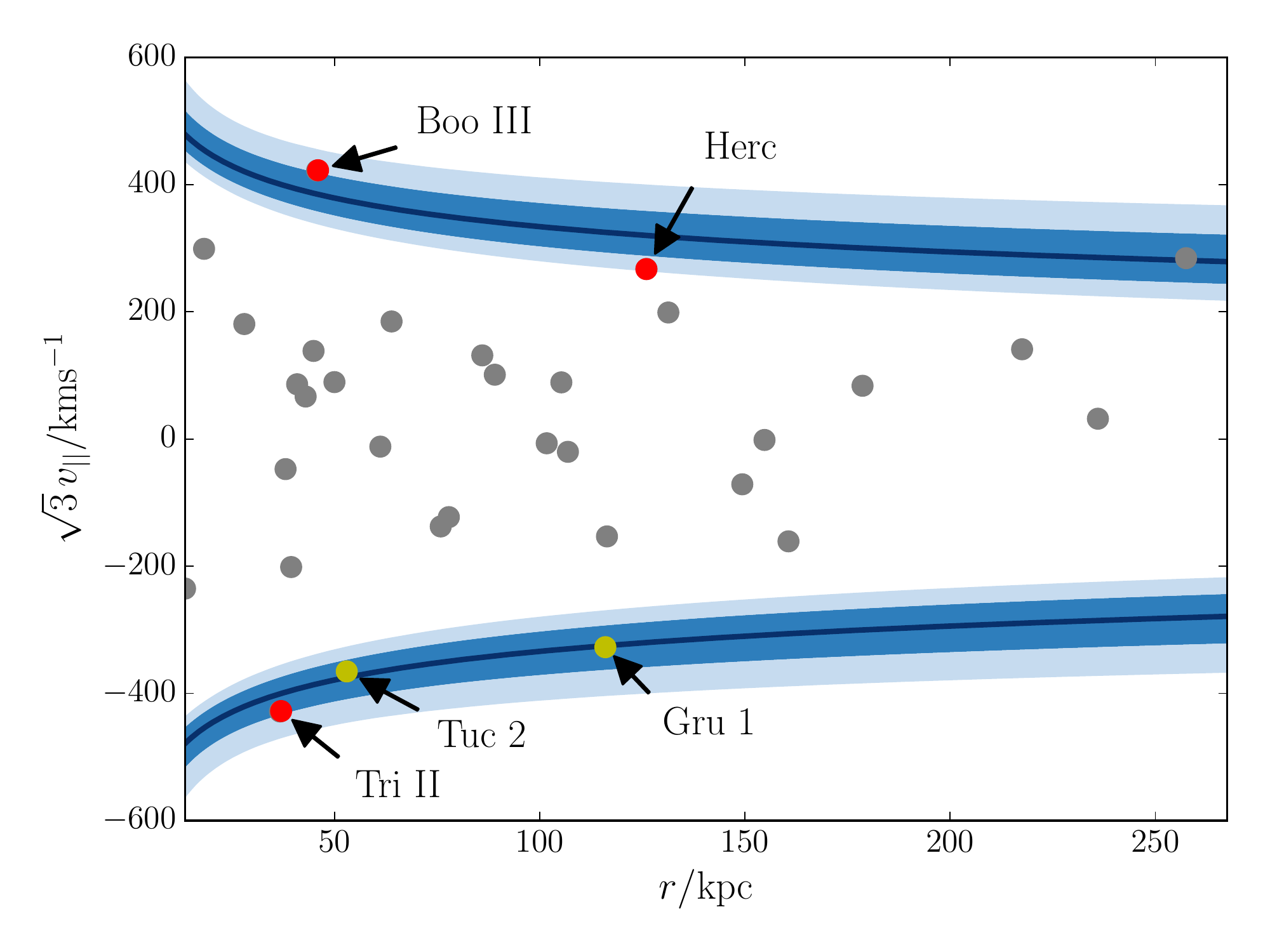}\\ 
\caption{The distribution of dwarf galaxies around the Milky Way in the $r-\sqrt{3}\,\vlos$ plane. 
The blue bands are our inference on the escape speed as a function of radius. 
Radial velocities have been multiplied by $\sqrt{3}$, as is done in the literature to account for unknown tangential velocities. 
If the true speeds of Triangulum II, Tucana II, Grus 1, Bootes III and Hercules are close to $\sqrt{3}\,\vlos$, then they are likely to be unbound.}
\label{fig:sats}
\end{figure}

Figure \ref{fig:sats} shows the distribution of known Milky Way dwarf galaxies in the $r-\sqrt{3}\,\vlos$ plane. 
It is typical in the literature to multiply the radial velocity by $\sqrt{3}$ as a crude way of accounting for unknown tangential velocities. 
We see that most of the dwarfs are enveloped by the escape speed curves, with a similar shape to the $r-\vlos$ distribution of stars (Figure \ref{fig:stars_bound}).
However, some of the dwarfs seem likely unbound based on our estimate of the escape speed. 
On the other hand, $\Lambda$CDM simulations predict that 99.9\% of subhalos should be bound to their hosts \citep{Bo13}. 
A reconciliation of these two statements is to conclude that the $\sqrt{3}\vlos$ approximation for the total speed of these dwarfs is likely unrealistic in these cases. 
Given our inference on the escape speed, the assumption that these objects are bound allows us to place constraints on their orbits. 
The red points in Figure \ref{fig:sats} are Bootes III \citep{Gr09}, Triangulum II \citep{La15} and Hercules \citep{Be07}, all of which are likely associated with the Milky Way and have large radial velocities. 
They are also at radii where our inference on the escape speed is trustworthy (although Hercules is a somewhat marginal case).  
The two yellow points are Tucana 2 and Grus 1 \citep{Ko15,Be15}, which are probably dwarfs of the Magellanic clouds \citep{Je16}. 
We do not seek to constrain the orbital properties of these dwarfs because of their more complex orbital histories. 
For the three Milky Way dwarfs, we would like to characterise their orbits through their pericenter radii $\rp$, apocenter radii $\ra$ and orbital eccentricities $\ecc$.

In order to compute posterior distributions on $\rp$, $\ra$ and $\ecc$, we first write:
\begin{equation}
p(X\,|\,v_r,r) = \int p(X\,|\,v_r,r,\Theta)\,p(\Theta\,|\,\mathrm{data})\,\intd \Theta,
\label{eq:dwarfpost}
\end{equation}
where $X = (\rp,\ra,\ecc)$ and $p(\Theta\,|\,\mathrm{data})$ is the posterior distribution on our model of the escape speed. 
We have written $\vlos \simeq v_r$, where $v_r$ is the velocity away from the centre of the Galaxy, because these objects are distant.  
We then have
\begin{align}
p(X\,|\,v_r,r,\,&\Theta) = \int p(X\,|\,v_r,v_T,r,\Theta)\,p(v_T\,|\,v_r,r,\Theta)\,\intd v_T, \nonumber \\
                     &= \int \delta\left[X - f_X(v_r,v_T,r,\Theta)\,\right]\,p(v_T\,|\,v_r,r,\Theta) \intd v_T,
\label{eq:deltafun}
\end{align}
where $v_T$ is the transverse velocity. 
In a spherical potential, the two velocities $(v_T,v_r)$ and radius $r$ completely specify the orbit, and hence $\rp$, $\ra$ and $\ecc$. 
This gives rise to the delta--function in Equation (\ref{eq:deltafun}), where $f_X$ represents the relation between $(v_r,v_T,r)$ and $X$. 
Finally, we need to specify $p(v_T\,|\,v_r,r,\Theta)$. 
To do this, we assume that the high speed dwarfs follow the same distribution function as the high speed stars, given by Equation (\ref{eq:lt90})
\begin{equation}
p(v_T\,|\,v_r,r,\Theta) \propto \left(\vesc(r) - \sqrt{v_T^2+v_r^2}\right)^k.
\label{eq:dwarfvT} 
\end{equation}
We make this assumption for simplicity, although \citet{Er16} note that the velocity distributions of subhaloes in the VLII simulations \citep{Di08} are not the same as those of the dark matter particles, and have velocity dispersions of size $160-200\kms$.
Measured velocity dispersions for Milky Way halo stars tend to be somewhat smaller than this, at $\sim 100-150\kms$ \citep{Ev16}.
Hence, for a given radial velocity, larger tangential velocities could be more probable than the estimate of Equation (\ref{eq:dwarfvT}), which in turn means that the orbit of the dwarf may be more eccentric than our simple calculation suggests.

Since $f_X$ is not analytic, we must solve for $\rp$, $\ra$ and $\ecc$ numerically. 
We do this by solving the equation 
\begin{equation}
E - \dfrac{L^2}{2r^2} - \Phi(r) = 0,
\label{eq:rperirapoeq}
\end{equation}
where $\Phi(r)$ is the gravitational potential implied by our model of the escape speed, $E$ is the orbital energy and $L$ is the total angular momentum.
This equation has two roots: $\rp$ and $\ra$. 
$\ecc$ is then given by
\begin{equation}
\ecc = \dfrac{\ra - \rp}{\ra + \rp}.
\end{equation}
In practice, we sample $p(X\,|\,v_r,r)$ in the following way. 
Given a set of parameters in our posterior samples for the SPL model, we first draw a tangential velocity from Equation (\ref{eq:dwarfvT}). 
For $k$, we use $k_\mathrm{MSTO}$, and we normalise the speed distribution between $v_r$ and $\vesc(r)$. 
Then, we solve Equation (\ref{eq:rperirapoeq}) for $\rp$, $\ra$ and $\ecc$ and store the result. 
This process is repeated for every set of model parameters in our posterior samples. 
The resulting histograms in $\rp$, $\ra$ and $\ecc$ are then faithful representations of $p(X\,|\,v_r,r)$.

The results of this procedure are shown in Figure \ref{fig:bootriherc}. 
All three dwarfs are expected to be on very eccentric orbits. 
Bootes III and Triangulum II have $\epsilon \simeq 0.95$, while Hercules has $\epsilon \simeq 0.8$, though with a somewhat broader distribution. 
This is aligned with our intuition: if the radial velocity alone is relatively close to the escape speed at the radius of the dwarf, then the tangential velocity cannot be large and hence the orbit must be eccentric. 
As a consequence, the dwarfs have large apocenters, $\ra \sim 100 - 300\kpc$ and considerably smaller pericenters $\rp \sim 10\kpc$ (although the posterior on the pericenter of Hercules is significantly less peaked than for Bootes III and Triangulum II).

\citet{Ku16} argued that the observed ellipticity of Hercules ($e \simeq 0.7$) and its large radial velocity are suggestive that it has `exploded' as a consequence of its last pericenter passage. 
Using N-body simulations, they arrived at an estimate of the orbital eccentricity $\ecc \simeq 0.95$, which is larger than our value but not in significant tension with it. 
Bootes III has a similar morphology ($e \simeq 0.5$) in keeping with the picture that these satellites are on orbits that will cause them to disrupt into streams after comparatively few orbital periods. 
Both Bootes II and Hercules have positive radial velocities: they are travelling away from the Galactic centre, suggesting that they have undergone at least one pericenter passage. 
Triangulum II, on the other hand, has a negative radial velocity and a relatively small ellipticity ($e \simeq 0.2$). 
Therefore, it is plausible that Triangulum II is on first infall and is about to undergo a large amount of disruption on its first pericenter passage.

\begin{figure*}
\includegraphics[width=2\columnwidth]{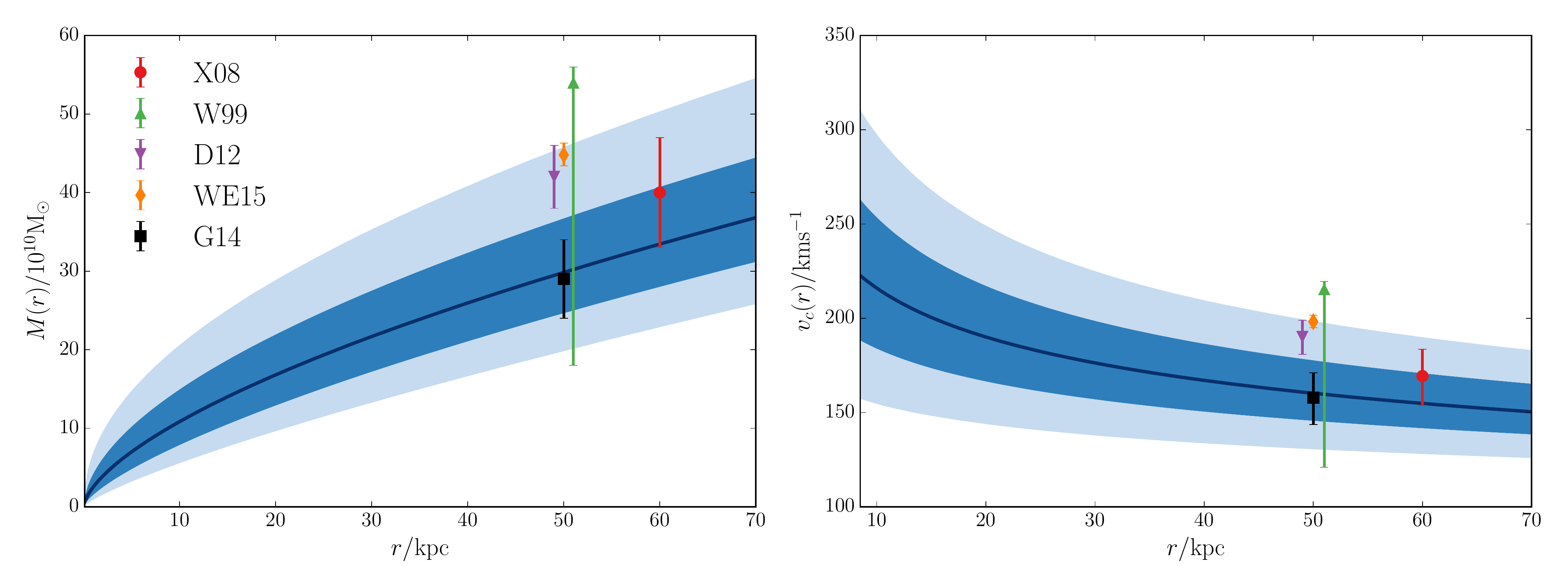}\\
\caption{Left: cumulative mass distribution within spherical shells, predicted by the SPL model. 
The 68\% (94\%) credible interval is shown as a dark (light) blue band around our median result. 
We predict a relatively light Milky Way within $50\kpc$, with $M(50\kpc)=29.8^{+6.9}_{-5.2} \times 10^{10}\mathrm{M}_\odot$. 
In both panels are the results from various other studies, see text for discussion.}
\label{fig:mass}
\end{figure*}

\subsection{Model performance and systematics}
\label{sec:model_performance}

We now seek to assess how well our model fits the data by performing posterior predictive checks. 
Since our model is only generative in radial velocities, and not in the positions of stars, we should compare the distribution in $\vlos$ of the data with our model. 
There is a subtlety in how this must be done, however. 
The velocity distribution of high-speed stars is position dependent in the model, owing to the spatial variation of the escape speed. 
The fastest star at a given radius is likely to be travelling slower than its counterparts at smaller radii, where the escape speed is larger. 
This effect means that we must take into account the number of stars that have been observed at each radius in the Galaxy for our comparison between model and data to be meaningful. 
We therefore write
\begin{eqnarray}
p(\vlos \,|\, \Theta) &=& \dfrac{\int p(\vlos \,|\, r, \Theta)\,p(r)\,\intd r}{\int p(\vlos | r, \Theta)\,p(r)\,\intd r\,\intd v} \nonumber \\
                  &=& \dfrac{\int (\vesc(r) - \vlos)^{k+1}\,p(r)\,\intd r}{\int (k+2)^{-1}(\vesc(r) - \vmin)^{k+2}\,p(r)\,\intd r\,},
\label{eq:ppc}
\end{eqnarray}
where $p(r)$ is the probability of observing a star at radius $r$. 
We approximate $p(r)$ by binning the mother sample of each tracer in radius and fitting a cubic spline to the resulting histogram.  
The two integrals in Equation (\ref{eq:ppc}) are then computed using 12 point Gauss-Legendre quadrature.  
We then draw samples from the posterior predictive distribution
\begin{equation}
p(\vlos\,|\,\mathrm{data}) = \int p(\vlos\,|\,\Theta)\,p(\Theta\,|\,\mathrm{data})\,\intd \Theta,
\end{equation}
which is the predicted distribution of $\vlos$ given the knowledge we have gained by analysing our SDSS sample.  
To sample this distribution, we first take the discrete samples from the posterior generated by our MCMC runs. 
Then, for each set of parameters $\Theta$, we draw a sample from Equation (\ref{eq:ppc}) that is the same size as the data, giving us many replicated data sets \citep{Ge13}. 
If the model is performing well, a typical replicated data set should look like the SDSS sample. 
In order to sample from Equation (\ref{eq:ppc}), we use inverse transform sampling, where we numerically compute the CDF
\begin{equation}
F(\vlos \,|\, \Theta) = \int\limits_{v_\mathrm{min}}^{\vlos} p(\vlos'\,|\,\Theta)\,\intd \vlos'
\end{equation}
on a grid of points in $\vlos$, and compute the inverse function $F^{-1}$ as a cubic spline. 
Finally, we draw a set of points $u$, uniformly distributed between 0 and 1, and compute the corresponding velocities via $\vlos = F^{-1}(u)$.

Figure \ref{fig:ppc} shows the comparison of the posterior predictive distributions with the data for our MSTO, K-giant and BHB samples. 
For each tracer, we constructed a histogram in radial velocity, and show the number of counts as a black point at each bin centroid. 
The median number of counts in each bin from our replicated data sets is shown as a solid line in each panel.
The 68\% and 95\% intervals are shown as bands around the median, and the final band shows the full extent of the number of counts.  
Our model reproduces the data very well over a range of $\sim 2.5$ orders of magnitude in the number of counts for the MSTO sample. 
Similarly good agreement is seen for our other two tracers.

Besides verifying that the model is a good representation of the data, we also seek to understand some of the possible sources of systematic uncertainty.
We investigated three possibilities: our choice of the local standard of rest $v_\mathrm{LSR}$, our choice of the cut velocity $\vmin$, and inconsistencies between the different tracer groups.
In the analysis presented in the rest of the paper, we assumed $v_\mathrm{LSR} = 240\kms$. 
In order to test the influence that this assumption has on our inference, we re--ran all of our analyses with a lower value of $v_\mathrm{LSR} = 220\kms$.
The local escape speed is then inferred to be $542^{+56}_{-37}\kms$, which is consistent with our previous analysis (note that different stars will enter our high-speed sample when a different value of $v_\mathrm{LSR}$ is used). 
The same is true for the rest of the model parameters.

We chose $\vmin=200\kms$ because we were not particularly concerned about contamination from disc stars. 
In order to check the sensitivity of our work to this value, we re--ran the analysis with $\vmin = 250\kms$. 
This results in a significantly smaller sample of 644 stars (539 MSTO, 99 K-giants, 6 BHBs). 
We thus expect much larger uncertainties on all of our model parameters. 
This is indeed the case, and we find that the values of $k$ for each of the tracers are poorly constrained compared to our full analysis, leading to a worse determination of the local escape speed. 
This is because the degeneracy between $k$ and $\vesc(R_\odot)$ is not broken as effectively by these data, which pushes up our estimates of $k$ and $\vesc(R_\odot)$, as predicted by S07. 
Specifically, we find $\vesc(R_\odot) = 617^{+77}_{-84}\kms$.  
This value is nonetheless consistent with our full analysis, due to the inflated uncertainties. 
Correspondingly, the inferred values of $k$ are $\sim 6$, with uncertainties $\sim 2$, which again are consistent with our previous estimates, but systematically higher. 
Given how well our full model represents the data, we would suggest that the speed distribution of halo stars does not significantly deviate from a power law at speeds $>200\kms$, thus vindicating our choice of $\vmin = 200\kms$. 
Larger choices of $\vmin$ should only be necessary in circumstances where disc contamination is a more serious concern. 
If this is the case, then our checks with $\vmin = 250\kms$ imply that a prior on the value of $k$ is probably necessary.

Our final check was to understand whether there is any tension between the different tracer samples. 
We ran our analysis again using only the MSTO sample, and then with the K-giants alone. 
We did not perform a run with the very small sample of BHBs. 
We found no tension whatsoever between the results from the MSTO only run and the K-giant only run. 
The K-giants favour a marginally larger value of $\alpha$, but the difference is not marked: the median values of $\alpha$ for each run lie comfortably within the 68\% credible regions of the other run.

\section{Conclusions}

\begin{figure*}
\includegraphics[width=2\columnwidth]{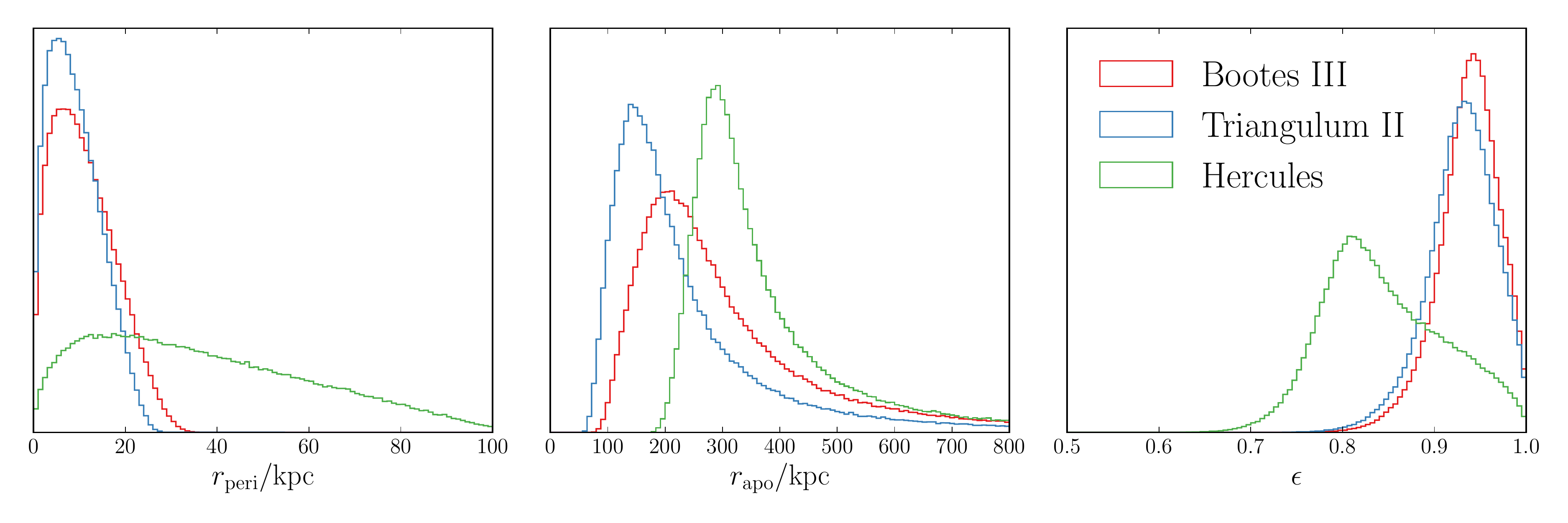}\\
\caption{Posterior distributions on orbital pericenter $r_\mathrm{peri}$, apocenter $r_\mathrm{apo}$ and eccentricity $\epsilon$ for Bootes III, Triangulum II and Hercules. 
If the dwarfs are bound, they must be on very eccentric orbits.}
\label{fig:bootriherc}
\end{figure*}

The main achievement of this paper is the measurement of the escape speed and its variation over Galactocentric radii between $\sim 8$ and $50$ kpc.  
This was done using a variety of tracer populations -- main sequence turn-off stars (MSTOs), blue horizontal branch stars (BHBs) and K giants -- extracted from the Sloan Digital Sky Survey. 
We find that the local escape speed is $521^{+46}_{-30}\kms$ in good agreement with the studies by \cite{Sm07} and \cite{Pi14} using data from the RAVE survey.

Our sample extends out to Galactocentric radii $\sim 50$ kpc, so we can track the gradient in the escape speed with radius for the first time. 
At $50\kpc$, the escape speed has fallen to a value of $379^{+34}_{-28}\kms$, indicative of a rapid power law decline ($\alpha = 0.37\pm0.09$).
This suggests that the total mass contained within $50\kpc$ is $29^{+7}_{-5}\times10^{10}\Msun$, implying a relatively light dark halo for the Milky Way. 
Our mass is pleasingly consistent with the work of \citet{Gi14}, who modelled the disruption of the Sagittarius dwarf galaxy, and therefore made very different assumptions in the analysis.

Perhaps the most striking thing about this method is its simplicity. 
We have demonstrated that modelling the tail of the velocity distribution can recover results in excellent agreement with other, much more elaborate, methods. 
In particular, there is no evidence of any bias as the values of the escape velocity and circular speed are in good agreement with other methods.  
As the size of data sets becomes huge, the computational cost of any algorithm will become an important yardstick. 
Our method is computationally much cheaper than full distribution function approaches, and will be able to rapidly provide robust results when faced with the output of enormous surveys.

The advent of the second Gaia release in 2017 and the first results from the WEAVE spectroscopic survey in 2018 will provide proper motions and radial velocities for massive samples of halo stars. 
The computational challenges of modelling such data sets are severe. 
We suggest that analysis of the high velocity tails may be the optimum method to quickly extract information on the potential of the Galaxy.  
Current data sets preclude a stringent measurement on the shape of the halo using the variation in escape speed with position, but we anticipate that this too will become possible in the very near future.

As well as allowing us to infer the escape speed and Galactic potential, the tail of the velocity distribution also provides information about the formation history of the Galaxy.
If the stellar halo is formed by some number of disrupted dwarf galaxies, then the power law index $k$ encodes their accretion times: we expect $k \equiv k(\tau)$, where $\tau$ is the time since infall of the dwarf progenitor of a given halo star.
We are investigating how this information can be extracted from cosmological simulations, in order to develop the machinery required to constrain $k(\tau)$. 
This method will turn the high velocity tail into a looking glass into the Milky Way's past.

As an application, we have used our model of the escape speed to study the orbital properties of the extreme Galactic satellites, Hercules, Bootes III and Triangulum II. 
All three must have highly eccentric orbits, with Bootes III and Triangulum II possessing an eccentricity $\epsilon \simeq 0.95$, and Hercules
$\epsilon \simeq 0.8$. 
The dwarfs must have large apocenters, $\ra \sim 100 - 300\kpc$ and small pericenters $\rp \sim 10\kpc$. 
Given that Hercules and Bootes III are now moving radially outwards, this implies that they must have at least one pericentric passage already and so we expect their morphology to be elongated and disrupted, as is the case. 
Triangulum II is moving inwards, and so must be on first infall, which is consistent with its roundish shape and ellipticity of 0.2.

\begin{figure*}
\includegraphics[width=1.5\columnwidth]{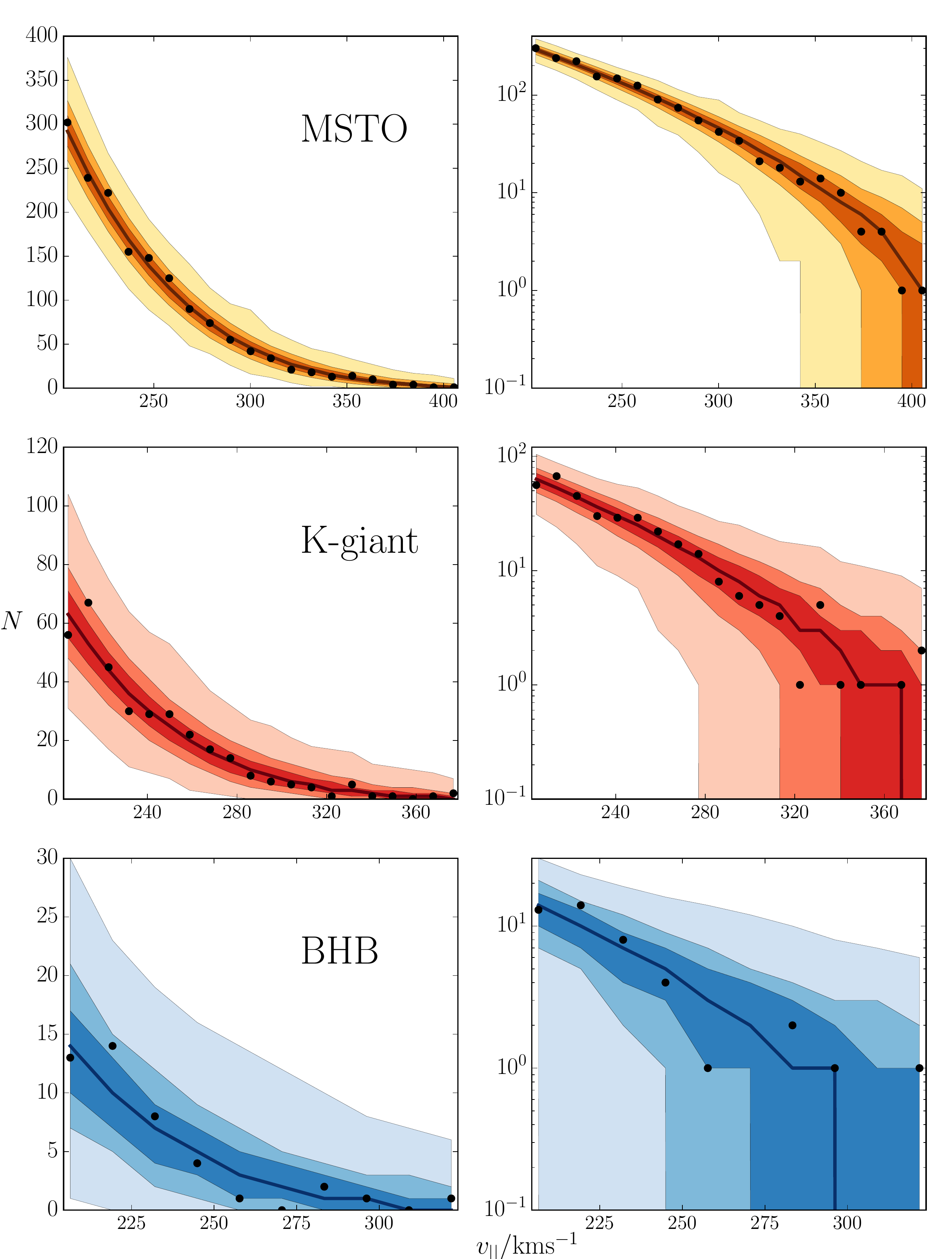}\\
\caption{Posterior predictive checks of the SPL model. 
The black points are the number of stars observed in bins of $\vlos$. 
The solid lines are the median counts from our replicated data sets. 
The coloured bands around the median line are the 68\%, 95\% intervals and full range of the counts in the replicated data sets. 
Our model produces replicated data that resemble the SDSS sample.}
\label{fig:ppc}
\end{figure*}

\section*{Acknowledgments}
AAW acknowledges the support of STFC. 
We thank Sergey Koposov for plenty of helpful advice, and Prashin Jethwa, Jason Sanders, Kathryn Johnston, Jorge Pe\~{n}arrubia and Gabriel Torrealba for useful conversations. 
This project was developed in part at the 2016 NYC Gaia Sprint, hosted by the Center for Computational Astrophysics at the Simons Foundation in New York City.

\bibliography{ontherun}

\begin{thebibliography}{}

\bibitem[\protect\citeauthoryear{{Ahn}, {Alexandroff}, {Allende Prieto},
  {Anderson}, {Anderton}, {Andrews}, {Aubourg}, {Bailey}, {Balbinot}, {Barnes}
  \& et al.}{{Ahn} et~al.}{2012}]{Ah12}
{Ahn} C.~P.,  {Alexandroff} R.,  {Allende Prieto} C.,  {Anderson} S.~F.,
  {Anderton} T.,  {Andrews} B.~H.,  {Aubourg} {\'E}.,  {Bailey} S.,  {Balbinot}
  E.,  {Barnes} R.,    et al. 2012, \apjs, 203, 21

\bibitem[\protect\citeauthoryear{{Alexander}}{{Alexander}}{1982}]{Al82}
{Alexander} J.~B.,  1982, \mnras, 201, 579

\bibitem[\protect\citeauthoryear{{Bechtol} et~al.,}{{Bechtol}
  et~al.}{2015}]{Be15}
{Bechtol} K.,  et~al., 2015, \apj, 807, 50

\bibitem[\protect\citeauthoryear{{Belokurov} et~al.,}{{Belokurov}
  et~al.}{2007}]{Be07}
{Belokurov} V.,  et~al., 2007, \apj, 654, 897

\bibitem[\protect\citeauthoryear{{Belokurov}, {Koposov}, {Evans},
  {Pe{\~n}arrubia}, {Irwin}, {Smith}, {Lewis}, {Gieles}, {Wilkinson},
  {Gilmore}, {Olszewski} \& {Niederste-Ostholt}}{{Belokurov}
  et~al.}{2014}]{Be14}
{Belokurov} V.,  {Koposov} S.~E.,  {Evans} N.~W.,  {Pe{\~n}arrubia} J.,
  {Irwin} M.~J.,  {Smith} M.~C.,  {Lewis} G.~F.,  {Gieles} M.,  {Wilkinson}
  M.~I.,  {Gilmore} G.,  {Olszewski} E.~W.,    {Niederste-Ostholt} M.,  2014,
  \mnras, 437, 116

\bibitem[\protect\citeauthoryear{{Binney} \& {Tremaine}}{{Binney} \&
  {Tremaine}}{2008}]{BT08}
{Binney} J.,  {Tremaine} S.,  2008, {Galactic Dynamics: Second Edition}.
Princeton University Press

\bibitem[\protect\citeauthoryear{{Boubert} \& {Evans}}{{Boubert} \&
  {Evans}}{2016}]{Bou16}
{Boubert} D.,  {Evans} N.~W.,  2016, \apjl, 825, L6

\bibitem[\protect\citeauthoryear{{Bowden}, {Evans} \& {Williams}}{{Bowden}
  et~al.}{2016}]{Bo16}
{Bowden} A.,  {Evans} N.~W.,    {Williams} A.~A.,  2016, \mnras, 460, 329

\bibitem[\protect\citeauthoryear{{Boylan-Kolchin}, {Bullock}, {Sohn}, {Besla}
  \& {van der Marel}}{{Boylan-Kolchin} et~al.}{2013}]{Bo13}
{Boylan-Kolchin} M.,  {Bullock} J.~S.,  {Sohn} S.~T.,  {Besla} G.,    {van der
  Marel} R.~P.,  2013, \apj, 768, 140

\bibitem[\protect\citeauthoryear{{Brown}}{{Brown}}{2015}]{Br15}
{Brown} W.~R.,  2015, \araa, 53, 15

\bibitem[\protect\citeauthoryear{{Caldwell} \& {Ostriker}}{{Caldwell} \&
  {Ostriker}}{1981}]{Ca81}
{Caldwell} J.~A.~R.,  {Ostriker} J.~P.,  1981, \apj, 251, 61

\bibitem[\protect\citeauthoryear{{Deason}, {Belokurov} \& {Evans}}{{Deason}
  et~al.}{2011}]{De11}
{Deason} A.~J.,  {Belokurov} V.,    {Evans} N.~W.,  2011, \mnras, 416, 2903

\bibitem[\protect\citeauthoryear{{Deason}, {Belokurov}, {Evans} \&
  {An}}{{Deason} et~al.}{2012}]{De12}
{Deason} A.~J.,  {Belokurov} V.,  {Evans} N.~W.,    {An} J.,  2012, \mnras,
  424, L44

\bibitem[\protect\citeauthoryear{{Diemand}, {Kuhlen}, {Madau}, {Zemp}, {Moore},
  {Potter} \& {Stadel}}{{Diemand} et~al.}{2008}]{Di08}
{Diemand} J.,  {Kuhlen} M.,  {Madau} P.,  {Zemp} M.,  {Moore} B.,  {Potter} D.,
     {Stadel} J.,  2008, \nat, 454, 735

\bibitem[\protect\citeauthoryear{{Erkal}, {Belokurov}, {Bovy} \&
  {Sanders}}{{Erkal} et~al.}{2016}]{Er16}
{Erkal} D.,  {Belokurov} V.,  {Bovy} J.,    {Sanders} J.~L.,  2016, \mnras,
  463, 102

\bibitem[\protect\citeauthoryear{{Evans}, {Sanders}, {Williams}, {An},
  {Lynden-Bell} \& {Dehnen}}{{Evans} et~al.}{2016}]{Ev16}
{Evans} N.~W.,  {Sanders} J.~L.,  {Williams} A.~A.,  {An} J.,  {Lynden-Bell}
  D.,    {Dehnen} W.,  2016, \mnras, 456, 4506

\bibitem[\protect\citeauthoryear{{Foreman-Mackey}, {Hogg}, {Lang} \&
  {Goodman}}{{Foreman-Mackey} et~al.}{2013}]{emcee}
{Foreman-Mackey} D.,  {Hogg} D.~W.,  {Lang} D.,    {Goodman} J.,  2013, \pasp,
  125, 306

\bibitem[\protect\citeauthoryear{Gelman, Carlin, Stern, Dunson, Vehtari \&
  Rubin}{Gelman et~al.}{2013}]{Ge13}
Gelman A.,  Carlin J.,  Stern H.,  Dunson D.,  Vehtari A.,    Rubin D.,  2013,
  Bayesian Data Analysis, Third Edition.
Chapman \& Hall/CRC Texts in Statistical Science, Taylor \& Francis

\bibitem[\protect\citeauthoryear{{Gibbons}, {Belokurov} \& {Evans}}{{Gibbons}
  et~al.}{2014}]{Gi14}
{Gibbons} S.~L.~J.,  {Belokurov} V.,    {Evans} N.~W.,  2014, \mnras, 445, 3788

\bibitem[\protect\citeauthoryear{{Goodman} \& {Weare}}{{Goodman} \&
  {Weare}}{2010}]{Go10}
{Goodman} J.,  {Weare} J.,  2010, Commun. Appl. Math. Comput. Sci., 5, 65

\bibitem[\protect\citeauthoryear{{Grillmair}}{{Grillmair}}{2009}]{Gr09}
{Grillmair} C.~J.,  2009, \apj, 693, 1118

\bibitem[\protect\citeauthoryear{{Ivezi{\'c}} et~al.,}{{Ivezi{\'c}}
  et~al.}{2008}]{Iv08}
{Ivezi{\'c}} {\v Z}.,  et~al., 2008, \apj, 684, 287

\bibitem[\protect\citeauthoryear{{Jethwa}, {Erkal} \& {Belokurov}}{{Jethwa}
  et~al.}{2016}]{Je16}
{Jethwa} P.,  {Erkal} D.,    {Belokurov} V.,  2016, \mnras, 461, 2212

\bibitem[\protect\citeauthoryear{{Koposov}, {Belokurov}, {Torrealba} \&
  {Evans}}{{Koposov} et~al.}{2015}]{Ko15}
{Koposov} S.~E.,  {Belokurov} V.,  {Torrealba} G.,    {Evans} N.~W.,  2015,
  \apj, 805, 130

\bibitem[\protect\citeauthoryear{{Kordopatis} et~al.,}{{Kordopatis}
  et~al.}{2013}]{Ko13}
{Kordopatis} G.,  et~al., 2013, \aj, 146, 134

\bibitem[\protect\citeauthoryear{{K{\"u}pper}, {Johnston}, {Mieske}, {Collins}
  \& {Tollerud}}{{K{\"u}pper} et~al.}{2016}]{Ku16}
{K{\"u}pper} A.~H.~W.,  {Johnston} K.~V.,  {Mieske} S.,  {Collins} M.~L.~M.,
  {Tollerud} E.~J.,  2016, ArXiv e-prints

\bibitem[\protect\citeauthoryear{{Laevens} et~al.,}{{Laevens}
  et~al.}{2015}]{La15}
{Laevens} B.~P.~M.,  et~al., 2015, \apjl, 802, L18

\bibitem[\protect\citeauthoryear{{Leonard} \& {Tremaine}}{{Leonard} \&
  {Tremaine}}{1990}]{Le90}
{Leonard} P.~J.~T.,  {Tremaine} S.,  1990, \apj, 353, 486

\bibitem[\protect\citeauthoryear{{McMillan} \& {Binney}}{{McMillan} \&
  {Binney}}{2013}]{Mc13}
{McMillan} P.~J.,  {Binney} J.~J.,  2013, \mnras, 433, 1411

\bibitem[\protect\citeauthoryear{{Piffl} et~al.,}{{Piffl}  et~al.}{2014}]{Pi14}
{Piffl} T.,  et~al., 2014, \aap, 562, A91

\bibitem[\protect\citeauthoryear{{Scannapieco}, {White}, {Springel} \&
  {Tissera}}{{Scannapieco} et~al.}{2009}]{Sc09}
{Scannapieco} C.,  {White} S.~D.~M.,  {Springel} V.,    {Tissera} P.~B.,  2009,
  \mnras, 396, 696

\bibitem[\protect\citeauthoryear{{Sch{\"o}nrich}, {Binney} \&
  {Dehnen}}{{Sch{\"o}nrich} et~al.}{2010}]{Sc10}
{Sch{\"o}nrich} R.,  {Binney} J.,    {Dehnen} W.,  2010, \mnras, 403, 1829

\bibitem[\protect\citeauthoryear{{Smith} et~al.,}{{Smith}  et~al.}{2007}]{Sm07}
{Smith} M.~C.,  et~al., 2007, \mnras, 379, 755

\bibitem[\protect\citeauthoryear{{Spitzer} Jr. \& {Shapiro}}{{Spitzer} \&
  {Shapiro}}{1972}]{Sp72}
{Spitzer} Jr. L.,  {Shapiro} S.~L.,  1972, \apj, 173, 529

\bibitem[\protect\citeauthoryear{{Theuns} \& {Warren}}{{Theuns} \&
  {Warren}}{1997}]{Th97}
{Theuns} T.,  {Warren} S.~J.,  1997, \mnras, 284, L11

\bibitem[\protect\citeauthoryear{{Vibert Douglas}}{{Vibert
  Douglas}}{1956}]{Vi56}
{Vibert Douglas} A.,  1956, {Arthur Stanley Eddington}.
Cambridge University Press

\bibitem[\protect\citeauthoryear{{Wang}, {Han}, {Cooper}, {Cole}, {Frenk} \&
  {Lowing}}{{Wang} et~al.}{2015}]{Wa15}
{Wang} W.,  {Han} J.,  {Cooper} A.~P.,  {Cole} S.,  {Frenk} C.,    {Lowing} B.,
   2015, \mnras, 453, 377

\bibitem[\protect\citeauthoryear{{Wilkinson} \& {Evans}}{{Wilkinson} \&
  {Evans}}{1999}]{Wi99}
{Wilkinson} M.~I.,  {Evans} N.~W.,  1999, \mnras, 310, 645

\bibitem[\protect\citeauthoryear{{Williams} \& {Evans}}{{Williams} \&
  {Evans}}{2015}]{Wi15}
{Williams} A.~A.,  {Evans} N.~W.,  2015, \mnras, 454, 698

\bibitem[\protect\citeauthoryear{{Xue}, {Ma}, {Rix}, {Morrison}, {Harding},
  {Beers}, {Ivans}, {Jacobson}, {Johnson}, {Lee}, {Lucatello}, {Rockosi},
  {Sobeck}, {Yanny}, {Zhao} \& {Allende Prieto}}{{Xue} et~al.}{2014}]{Xu14}
{Xue} X.-X.,  {Ma} Z.,  {Rix} H.-W.,  {Morrison} H.~L.,  {Harding} P.,  {Beers}
  T.~C.,  {Ivans} I.~I.,  {Jacobson} H.~R.,  {Johnson} J.,  {Lee} Y.~S.,
  {Lucatello} S.,  {Rockosi} C.~M.,  {Sobeck} J.~S.,  {Yanny} B.,  {Zhao} G.,
   {Allende Prieto} C.,  2014, \apj, 784, 170

\bibitem[\protect\citeauthoryear{{Xue}, {Rix}, {Zhao}, {Re Fiorentin}, {Naab},
  {Steinmetz}, {van den Bosch}, {Beers}, {Lee}, {Bell}, {Rockosi}, {Yanny},
  {Newberg}, {Wilhelm}, {Kang}, {Smith} \& {Schneider}}{{Xue}
  et~al.}{2008}]{Xu08}
{Xue} X.~X.,  {Rix} H.~W.,  {Zhao} G.,  {Re Fiorentin} P.,  {Naab} T.,
  {Steinmetz} M.,  {van den Bosch} F.~C.,  {Beers} T.~C.,  {Lee} Y.~S.,  {Bell}
  E.~F.,  {Rockosi} C.,  {Yanny} B.,  {Newberg} H.,  {Wilhelm} R.,  {Kang} X.,
  {Smith} M.~C.,    {Schneider} D.~P.,  2008, \apj, 684, 1143

\bibitem[\protect\citeauthoryear{{Yanny} et~al.,}{{Yanny}  et~al.}{2009}]{Ya09}
{Yanny} B.,  et~al., 2009, \aj, 137, 4377

\end{thebibliography}
\bibliographystyle{mn2e}

\appendix

\section{SQL queries}
\label{sec:sql}

Here we give the {\sc SQL} queries that we used to obtain the MSTO and BHB samples.

\begin{enumerate}
\item MSTO query:
\begin{verbatim}
SELECT *

FROM sdssdr9.specphotoall AS spa,
     sdssdr9.sppparams AS spp

WHERE spp.specobjid=spa.specobjid
AND spp.scienceprimary=1 
AND spa.class=`STAR' 
AND spa.extinction_r<0.3
AND spa.dered_g-spa.dered_r BETWEEN 0.2 AND 0.6 
AND spa.dered_r BETWEEN 14.5 AND 20. 
AND spp.fehadop BETWEEN -4. AND -0.9 
AND spp.loggadop BETWEEN 3.5 AND 4. 
AND spp.teffadop BETWEEN 4500. AND 8000.
AND spa.psfmagerr_g BETWEEN 0. AND 0.04
AND spa.psfmagerr_r BETWEEN 0. AND 0.04
AND spa.psfmagerr_i BETWEEN 0. AND 0.04
AND spa.fehadopunc < 0.1 
AND (spp.zwarning=0 OR spp.zwarning=16)
AND spp.snr > 20.
\end{verbatim}
When we apply our extra morphological cut, the condition
\begin{verbatim}
spa.TYPE = 6
\end{verbatim}
is included in the above query.
\item BHB query:
\begin{verbatim}
SELECT * 

FROM sdssdr9.specphotoall AS spa,
     sdssdr9.sppparams AS spp
     
WHERE spp.specobjid=spa.specobjid
AND spp.scienceprimary=1
AND spa.class=`STAR'
AND spa.psfmag_g-spa.extinction_g-spa.psfmag_r
    +spa.extinction_r BETWEEN -0.25 AND 0.
AND spa.psfmag_u-spa.extinction_u-spa.psfmag_g
    +spa.extinction_g BETWEEN 0.9 AND 1.4
AND spp.fehadop BETWEEN -2. AND -1.
AND spp.loggadop BETWEEN 3. AND 3.5
AND spp.teffadop BETWEEN 8300. AND 9300.
AND (spp.zwarning=0 OR spp.zwarning=16)
AND spp.snr>20.
\end{verbatim}
\end{enumerate}

\section{Results with stricter morphological cuts}

Here we show the results when objects that have not been morphologically classified as stars are removed from the MSTO catalog.

\begin{figure*}
\includegraphics[width=2\columnwidth]{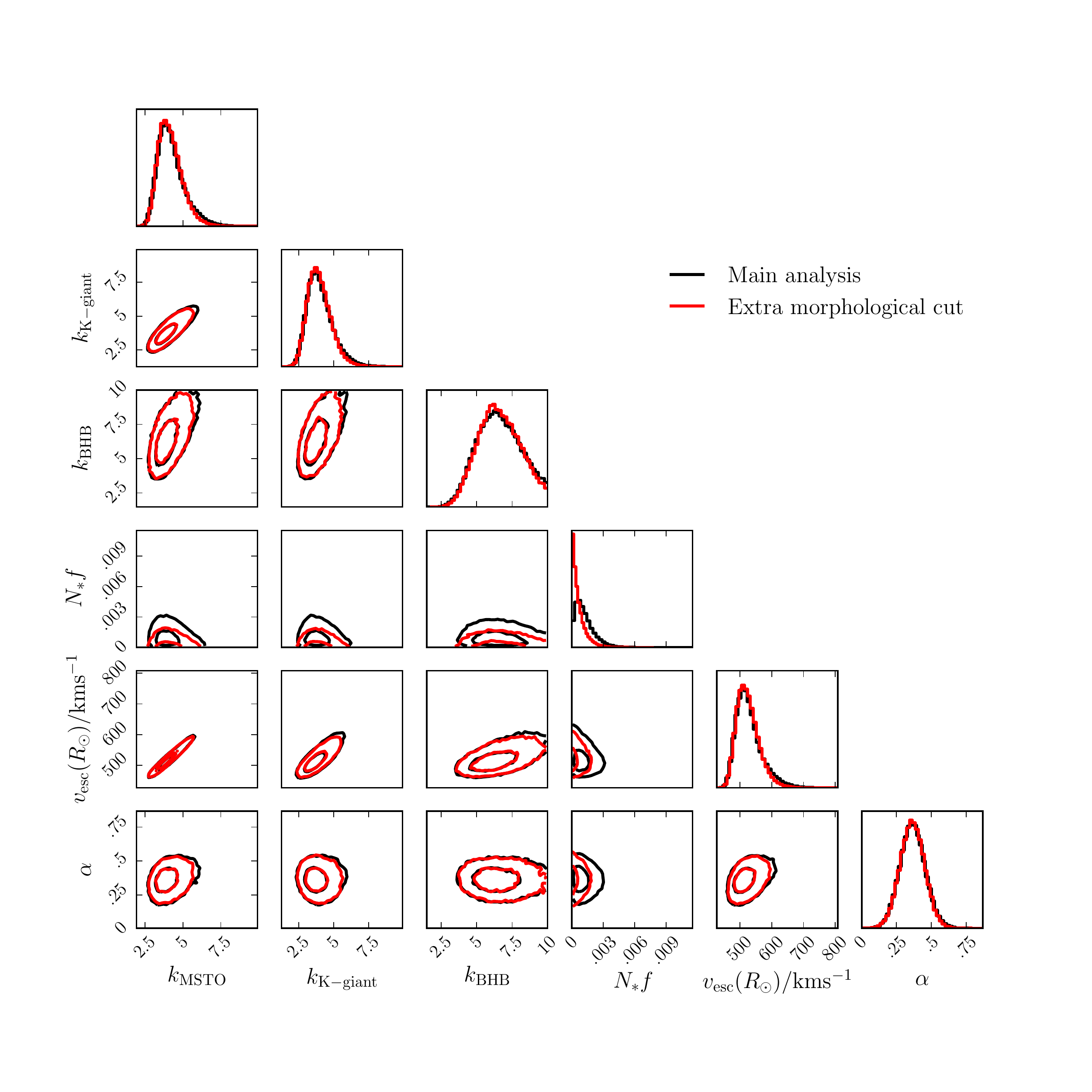}\\
\caption{One and two dimensional posterior distributions of two runs: our original SPL run, and the run where the stricter morphological cut discussed in Section \ref{sec:res} is applied.
The projections of the posterior distributions from the two runs are completely consistent with one another, save for the outlier fraction.
This is because, when the morphological cut is applied, the galaxy contaminant with a spurious radial velocity discussed in Section \ref{sec:res} is removed.}
\label{fig:morph_corner}
\end{figure*}

\label{sec:morph}

\end{document}